\documentclass[useAMS,usenatbib]{mn2e}

\usepackage{graphicx}
\usepackage{times}
\usepackage{natbib}
\usepackage{rotating}

\newcommand{\be}{\begin{equation}}
\newcommand{\ee}{\end{equation}}
\newcommand{\ba}{\begin{eqnarray}}
\newcommand{\ea}{\end{eqnarray}}
\newcommand{\brr}{\begin{array}}
\newcommand{\err}{\end{array}}
\newcommand{\bc}{\begin{center}}
\newcommand{\ec}{\end{center}}

\newcommand{\mincir}{\raise
  -2.truept\hbox{\rlap{\hbox{$\sim$}}\raise5.truept \hbox{$<$}\ }}
\newcommand{\magcir}{\raise
  -2.truept\hbox{\rlap{\hbox{$\sim$}}\raise5.truept \hbox{$>$}\ }}
\newcommand{\siml}{\raise
  -2.truept\hbox{\rlap{\hbox{$\sim$}}\raise5.truept \hbox{$<$}\ }}
\newcommand{\simg}{\raise
  -2.truept\hbox{\rlap{\hbox{$\sim$}}\raise5.truept \hbox{$>$}\ }}

\title[Dynamical difference between the cD galaxy and the stellar diffuse component in simulated galaxy clusters]
{Dynamical difference between the cD galaxy and the diffuse, stellar component in simulated galaxy clusters}
\author[K. Dolag, G. Murante, S. Borgani]
{K.~Dolag$^{1}$\thanks{E-mail: kdolag@mpa-garching.mpg.de},
G. Murante$^2$ and S. Borgani$^{3,4,5}$\\
$^1$ Max-Planck-Institut f\"ur Astrophysik, Karl-Schwarzschild Strasse
  1, Garching bei M\"unchen, Germany (kdolag@mpa-garching.mpg.de)\\
$^2$ INAF -- Astronomical Observatory of Torino, Str. Osservatorio
25,  I-10025, Pino Torinese, Torino, Italy\\
$^3$ Dipartimento di Astronomia dell'Universit\`a di Trieste, via
  Tiepolo 11, I-34131 Trieste, Italy (borgani@oats.inaf.it)\\
$^4$ INFN -- Istituto Nazionale di Fisica Nucleare, Trieste, Italy\\
$^5$ INAF -- Istituto Nazionale di Astrofisica, Trieste, Italy\\
}

\begin{document}

\date{Accepted ???. Received ???; in original form ???}

\pagerange{\pageref{firstpage}--\pageref{lastpage}} \pubyear{0000}

\maketitle

\label{firstpage}

\begin{abstract}
Member galaxies within galaxy clusters nowadays can be routinely identified in
cosmological, hydrodynamical simulations using methods based on identifying
self bound, locally over dense substructures. However, distinguishing the central
galaxy from the stellar diffuse component within clusters is notoriously
difficult, and in the center it is not even clear if two distinct stellar
populations exist. Here, after subtracting all member galaxies, we use the
velocity distribution of the remaining stars and detect two dynamically,
well-distinct stellar components within simulated galaxy clusters. These
differences in the dynamics can be used to apply an un-binding procedure which
leads to a spatial separation of the two components into a {\it cD} and a
diffuse stellar component ({\it DSC}). Applying our new algorithm to a
cosmological, hydrodynamical simulation we find that -- in line with previous
studies -- these two components have clearly distinguished spatial and
velocity distributions as well as different star formation histories. We show
that the {\it DSC} fraction -- which can broadly be associated with the
observed intra cluster light -- does not depend on the virial mass of the
galaxy cluster and is much more sensitive to the formation history of the
cluster. We conclude that the separation of the {\it cD} and the {\it DSC} in
simulations, based on our dynamical criteria, is more physically motivated
than current methods which depend on implicit assumptions on a length scale
associated with the {\it cD} galaxy and therefore represent a step forward in
understanding the different stellar components within galaxy clusters.
Our results also show the importance of analyzing the dynamics of the DSC to
characterize its properties and understand its origin.
\end{abstract}



\section{Introduction} \label{sec:intro}

The existence of a diffuse stellar component in groups and clusters of
galaxies is now well established. A diffuse intracluster light ({\it ICL})
has been observed both in local universe
\citep{Feldmeier04,Mihos05,Arnaboldi04,Gerhard05,Doherty09} and at
intermediate redshift
\citep{Gonzalez00,Feldmeier04b,Zibetti05,Krick07}.
Such {\it ICL} component is centrally concentrated, and typically amounts 
between $\approx 10$ \citep{Zibetti05} and $\approx 35$ \% \citep{Gonzalez07} of the
total stellar mass in clusters. There are some indications pointing
towards a dependence of the {\it ICL} fraction from the mass of the
clusters, going from $\approx 2$ \% at the scale of loose groups
\citep{Castro-Rodr03} to $\approx 5-10$ \% of Virgo cluster
\citep{Magda03,Feldmeier04b,Mihos05,2009arXiv0908.3848C} up to $10-20$ \% or higher in the
most massive clusters
\citep{Gonzalez00,Feldmeier02,GalYam03,Feldmeier04b,Krick06}, but
other studies found instead no significant dependence on the cluster
richness \citep{Zibetti05}.

{\it ICL} is usually detected using surface photometry in various bands,
sometimes stacking together observations of many clusters \citep[see
  e.g.][]{Zibetti05}. In this case, light coming from galaxies is masked away,
often using a fixed limit in surface brightness \citep[e.g., as
  in][]{Feldmeier04}. Single {\it ICL} stars can also be detected
\citep{Durrell02}, expecially by observing Intra-Cluster Planetary Nebulae
(ICPN) \citep{Magda96,Feldmeier04b}, which can be observed up to distances as
large as 100 Mpc \citep{Gerhard05}. On the one hand, the observational effort
needed to detect single stars makes it difficult to use them to assess general
properties of the {\it ICL} distribution, like the {\it ICL} fraction in
clusters. On the other hand, the observation of single stars make it possible
to gather information on the kinematic of such a component.  In particular,
extended halos of bright ellipticals often overlap spatially with stars which
are free-floating in the cluster gravitational potential. Efforts have been
made to disentangle these two components \citep[see e.g.][]{Doherty09}, but
they are often referred together as the {\it ICL}. In the following, we will
use {\it ICL} to indicate the observed diffuse stellar population, whose
dynamical properties are difficult to obtain, and {\it DSC} for the
corresponding stellar population produced in numerical simulation, whose
dynamics is known.

From a theoretical point of view, a diffuse stellar component ({\it DSC}) has
been studied in a cosmological Dark-Matter-only numerical simulation
\citep{Napolitano03}.  Later, for the first time \cite{2004ApJ...607L..83M}
(M04 hereafter) studied the properties of the {\it DSC} in a cosmological
simulation which included hydrodynamics and various astrophysical processes
such as radiative cooling of gas and (self-consistent) star
formation. \cite{Willman04} and \cite{SommerLarsen05} analyzed a {\it DSC} in
their resimulation of single galaxy clusters.

M04 analyzed a cosmological simulation of a box of 192 $h^{-1}$ Mpc,
containing $\approx 100$ galaxy cluster. Their main results were that: (i) a
{\it DSC} is clearly seen when a Sersic fit is performed on the bound, unbound
and total two-dimensional radial density profiles of the stellar population in
clusters; (ii) the {\it DSC} component is more centrally concentrated than the
stars bound in galaxies; (iii) on average, stars in the {\it DSC} are older
than those in galaxies; (iv) the {\it DSC} fraction increase with the cluster
mass.  \cite{Willman04} used one cluster re-simulation, having a mass of $1.2
\times 10^{15} $ M$_\odot$, and confirmed results (i) of M04. They found a
{\it DSC} fraction compatible with M04 clusters in the same mass range, if
slightly lower. \cite{SommerLarsen05} used one Virgo-like and one Coma-like
cluster re-simulation and confirmed results (i), (ii) and (iii) of M04. They
found similar {\it DSC} fraction, with their Coma-like cluster having an
higher fraction of {\it DSC} stars. Their average stellar age in {\it DSC} was
older than that found by M04.

Several mechanisms have been investigated to explain the formation and
evolution of a {\it DSC}: stripping and disruption of galaxies as they pass
through the central regions of relaxed clusters
\citep{Byrd90,Gnedin03}; stripping of stars from galaxies during the
initial formation of clusters \citep{Merritt84}; creation of stellar
haloes in galaxy groups, that later fall into massive clusters and
then become unbound \citep{Mihos04,Rudick06}; and stripping of stars
during high-speed galaxy encounters in the cluster environment
\citep{Moore96}. \citet{2007MNRAS.377....2M} (M07 hereafter) showed that the {\it DSC}
is produced during mergers in the formation history of the BCGs and of
other massive galaxies, and that it grows steadily since redshift
$z=1$, with no preferred epoch of formation. \citet{Monaco06}, using a
semi-analytical model of galaxy formation in clusters, showed that, in
the hierarchical model of structure formation, the BCG merging
activity alone between $z=1$ and $z=0$ is so intense that it is not
possible to simultaneously fit the bright end of the galaxy luminosity
function at both redshifts without assuming that a significant
fraction of stars goes into the {\it DSC} during such
mergers. 
We point out that observational indications have been found that such 
process is in fact in action in the Coma cluster \citep{2007A&A...468..815G}. Numerically, 
\cite{Stanghellini06} studied isolated, star-only dry merger
of elliptical galaxies and showed how in that case, up to $21$ 
the initial total stellar mass can become unbind in the process. 

One major problem with the analysis of {\it DSC} in numerical simulations is
its very identification. One possibility, used e.g. by \cite{Rudick06}, is to
build simulated surface brightness maps and detect an {\it ICL} component with
a procedure similar to that used with observational data.  This has the
advantage of being somewhat easier to compare with observation, at the cost of
losing the intrinsic advantage of simulations, that is, the knowledge of the
dynamics of each star particle. The other possibility is to attempt a
dynamical distinction between stars bound to galaxies and stars which are
free-floating in the group or cluster potential. This is what M04,
\cite{Willman04} and \cite{SommerLarsen05} did; the disadvantage is that such
a distinction may or may not coincide with the observational definition of
{\it ICL}. However, for doing this, it is
necessary to identify structures in the simulation (groups or clusters) and
substructures within them (the galaxies). Up to date, no unambiguous way to
perform such identification has been formulated.

The identification of {\it structures} is a relatively easy task. The most
commonly used algorithms are based upon two methods: the
Friends-of-Friends (FoF) approach \citep{Davis85,Frenk88} and the Spherical
Overdensity one (SO) \citep{LaceyCole94}.
In the former, all particles closer than a given distance $l$ 
(usually expressed in units of the mean inter-particle distance, typically 
$l\approx0.2$ is chosen) are grouped
together; then, all particles closer than $l$ to such particles are included
in the group, and so on. 
In the latter, one identifies density peaks in the particle distribution, and
puts spheres around them, choosing their radius so that they enclose a given
overdensity. 

The difficulty arises when {\it sub}structures inside DM halos must be
found. The density contrast of substructures against their parent halo density
is lower than the density contrast of structures against the
background, and this makes their identification more problematic. Even worse,
the gravitational potential of a structure has a clear zero-point, given
by the background density of the universe. So, it is well defined which
particles are bound to the structure and what are unbound. Inside a structure,
instead, such a definition is not clear: the background density of the
structure varies inside them with the distance from the center, the dynamical
state of the cluster, and the typical extend of the substructure
itself. Therefore, telling that a particle is gravitationally bound to a given
substructures involves a certain degree of arbitrariness, and an operational
definition must be used. 

A number of different {\it sub}halo finders now exist in the literature.  
Some are based
on pure geometrical measures, others construct a candidate structure and apply an unbinding
criteria, e.g. comparing the kinetic energy of a particle to its potential energy.
A non-exhaustive list includes: hierarchical {\small FOF} \citep{Klypin99},  Bound Density
Maxima \citep{Klypin99}, {\small DENMAX} \citep{Bert91,Gelb94}, {\small SKID}
\citep{Weinberg97,2001PhDT........21S}, local density percolation \citep{1995MNRAS.273..295V},
 {\small HOP} \citep{Eisenstein98}, {\small SUBFIND}
\citep{2001MNRAS.328..726S}, MHF \citep{Gill04}, {\small ADAPTAHOP} \citep{Aubert04}, 
{\small VOBOZ} \citep{Ney05}, {\small PSB} \citep{Kim06}, {\small 6D FOF} \cite{Diemand06},
{\small HSF} \citep{Macie09}. {\small AHF} \citep{2009ApJS..182..608K}. 
Some of these algorithms are based on the FoF scheme, and vary the $l$
parameter inside structures or apply the scheme to the full 6D phase space
(hierarchical {\small FOF}, local density percolation, {\small 6D FOF}); 
some use the SO scheme, and apply an unbinding
criterion ({\small BDM}, {\small MHF}, {\small AHF}); some use FoF to link particles 
to density maxima, and
apply an unbinding ({\small DENMAX}, {\small SKID}); some introduce geometrical constraints,
together with various unbinding procedure and phase space analysis ({\small HOP},
{\small SUBFIND}, {\small ADAPTAHOP}, {\small VOBOS}, {\small PSB}, {\small HFS}).
Comparisons between the performance of the different algorithms in detecting
substructures usually show a reasonable agreement between the different methods
as can be found for example in \citet{2009ApJS..182..608K} or \citet{Macie09}.
Already \citet{2001MNRAS.328..726S} compared {\small SUBFIND} and {\small FOF}, showing that the
former performs better in those cases where nearby two halos are
connected by a filaments. {\small FOF} usually joins together the two objects,
while {\small SUBFIND} can disentangle them. Such a behaviour is however common
to most of the subhalo finder listed above.

As far as the {\it DSC} detection is concerned, a further question arises. The 
{\it DSC} is composed by stars {\it bound to the gravitational potential of the cluster
  and not to any galaxy}, and it is centrally concentrated. Many of its
general properties, therefore, are primarily determined by the stars at the center of the cluster. It
is therefore important to correctly distinguish {\it DSC} star particles from {\it cD}
star particles. But such a distinction is often somewhat arbitrary, as in {\small SKID},
where the unbinding procedure is based upon the value of a parameter that
must be given externally and whose effect must be tested {\it a-posteriori}
(see M07). Other substructure finders, like {\small SUBFIND}, identifies all
substructures but the main one and link all remaining star, gas and DM particles to
the main subhalo, thus mixing {\it cD} and {\it DSC} stars.
It is not even clear if a distinction between {\it cD} and its extended stellar halo
is physically motivated or if it is all arbitrary. 

In this work, we show that such two components can be distinguished in the
velocity space. We use {\small SUBFIND} to identify substructures and disentangle
galaxies from the {\it cD} + {\it DSC} stars. We show that the velocity distribution
of the latters cannot be fit with a single Maxwellian distribution, while it
is well fitted by a double Maxwellian (the sum of two Maxwellian distribution
with different velocity dispersions).

We thus modify {\small SUBFIND} to perform an unbinding procedure on the 
{\it cD} + {\it DSC} stars. We
use the gravitational potential given by the matter contained in a sphere,
centered on the halo center, and we find a radius which divides the star
particles in two populations whose velocity distribution is separately fit by a
single Maxwellian. We vary the radius until the two resulting velocity dispersions
correspond to the two velocity dispersion given by the double Maxwellian fit of
the velocity distribution of the whole {\it cD} + {\it DSC} star population. We show that
this procedure is able to disentangle the central galaxy from its extended
stellar halo. We thus characterize the {\it DSC} through its phase-space distribution.

We then use our modified {\small SUBFIND} algorithm to repeat the M04 analysis of the
{\it DSC} applied to a constrained cosmological simulation of the local universe. We also
use two re-simulations of galaxy clusters, to test the behavior of our
modified {\small SUBFIND} when we increase resolution. 

\citet{2010arXiv1001.3018P} used different ways to define the {\it DSC} in their 
simulations, including the one we present here. They found a broad qualitative
agreement among the various methods. Here, we will focus on the
comparison of our modified {\small SUBFIND} algorithm with {\small SKID} on our
constrained cosmological simulation of the local universe, and show
that they are in broad agreement, but dynamically separating the {\it cD} and the {\it DSC}, 
a dependance of the {\it DSC} fraction on cluster mass is not found. 

The plan of the paper is as follows. Section \ref{sec:simgen} describes our simulations.
Section \ref {sec:dynamics} shows the existence of two dynamically distinct
stellar components, even at the center of galaxy clusters. Section
\ref{sec:iclunbinding} presents our scheme for disentangling such components. 
In Section \ref{sec:cDDSC} we apply our scheme to our local universe
simulation, also comparing the results we obtain on the {\it DSC} properties
with those obtained using {\small SKID}. In Section \ref{sec:conc} we give our
conclusions. In the Appendix we show a detailed comparison of the performance
of {\small SKID} and {\small SUBFIND} for two galaxy clusters.


\section{General setup} \label{sec:simgen}

\begin{figure*}
\includegraphics[width=0.85\textwidth]{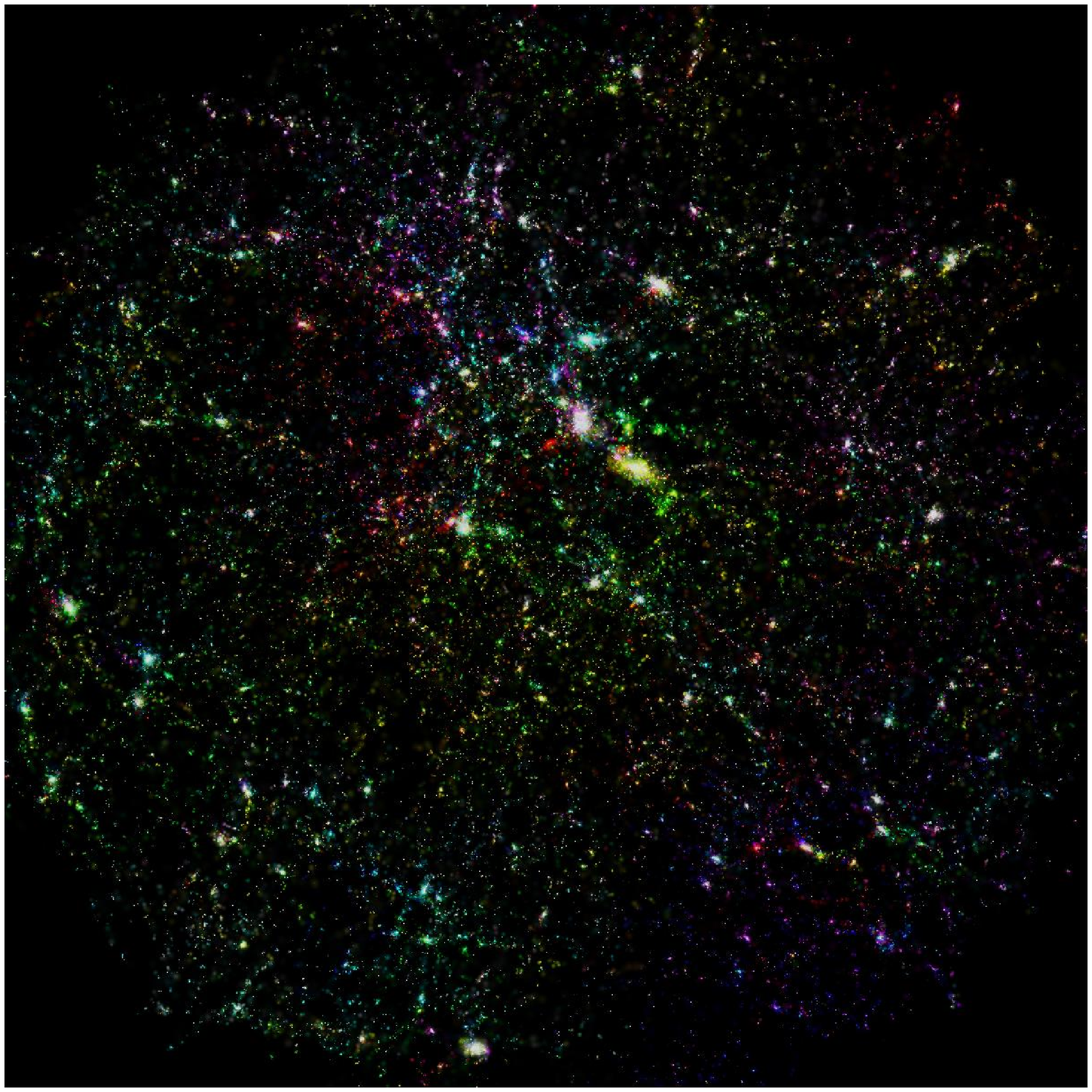}\\
\includegraphics[width=0.425\textwidth]{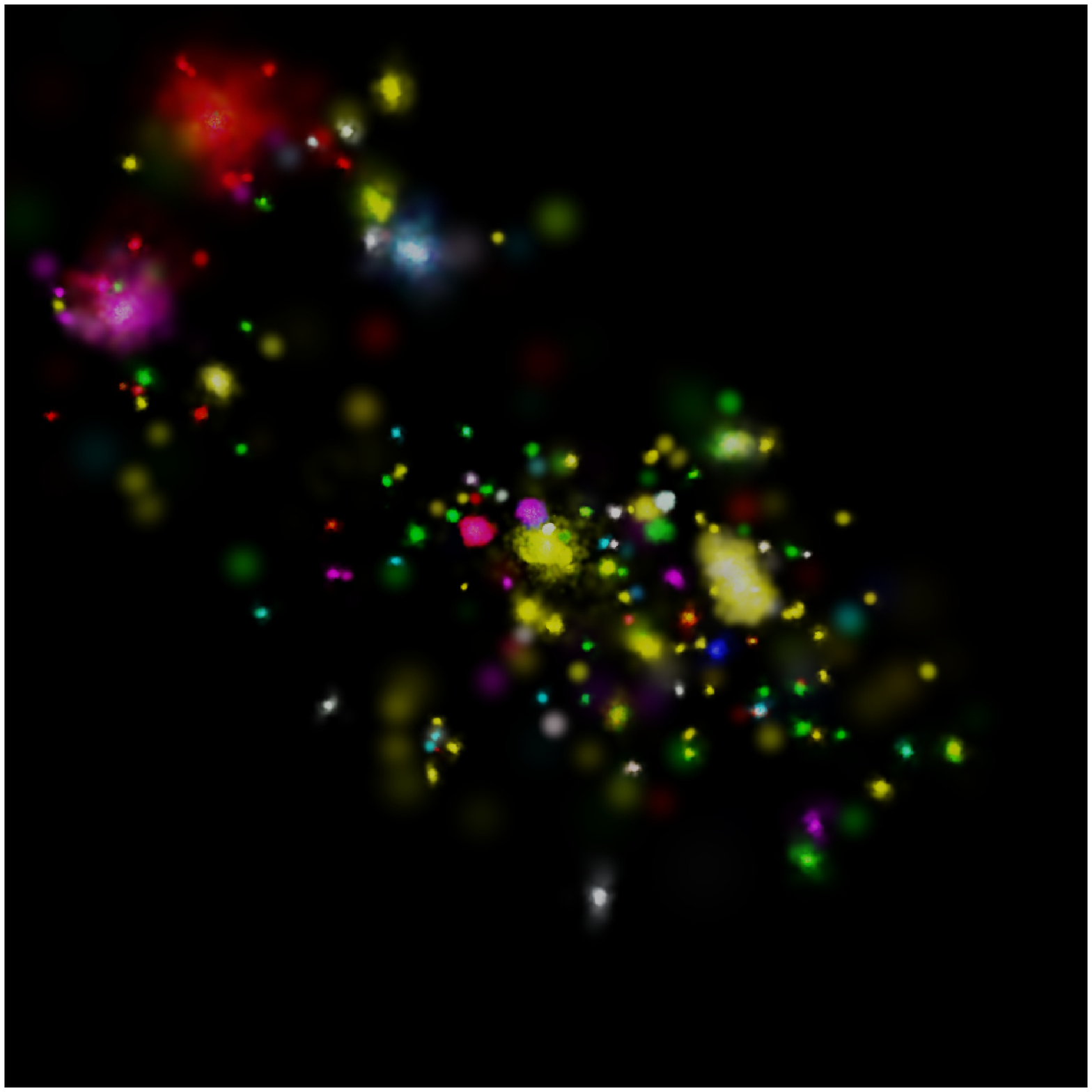}
\includegraphics[width=0.425\textwidth]{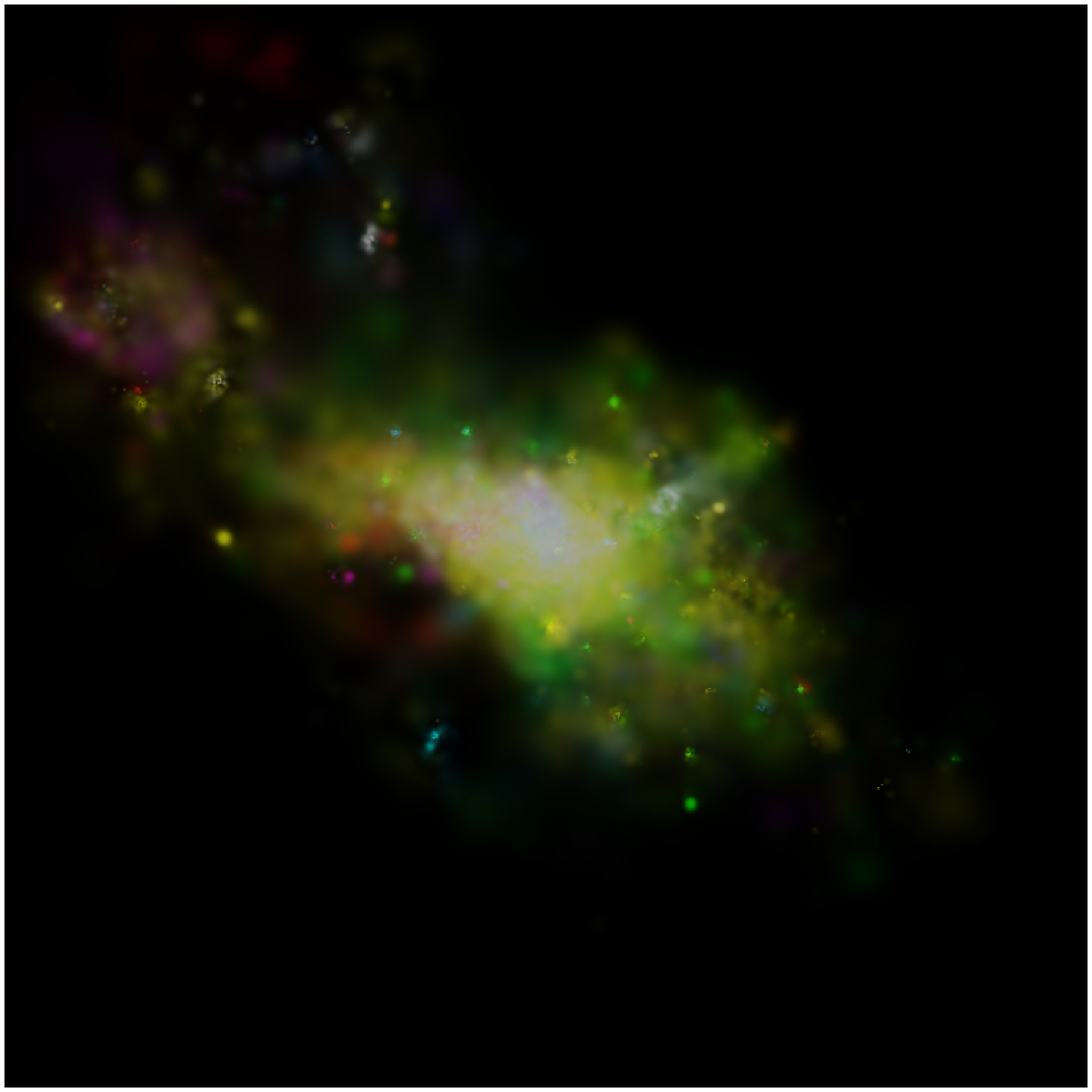}
\caption{Visualization of the stellar component using the 
the ray--tracing software {\small SPLOTCH}. The color composition reflect the
three dimensional velocity field (see text for details), whereas the intensity 
reflects the stellar density. The upper panel shows all stars in the high
resolution sphere. The lower left panel shows all stars in galaxies (including 
the {\it cD}) of the most massive cluster, whereas the lower right panel shows all 
stars not bound to individual galaxies.} \label{fig:stars_ray}
\end{figure*}

\subsection{Simulations} \label{sec:sim}
The results presented in this paper have been obtained using the final output
of a cosmological hydrodynamical simulation of the local universe. Our initial
conditions are similar to those adopted by \citet{Mathis:2002} in their study
(based on a pure N-body simulation) of structure formation in the local
universe.  The galaxy distribution in the IRAS 1.2-Jy galaxy survey is first
Gaussian smoothed on a scale of 7 Mpc and then linearly evolved back in time
up to $z=50$ following the method proposed by \cite{Kolatt:1996}. The
resulting field is then used as a Gaussian constraint \citep{Hoffman1991} for
an otherwise random realization of a flat $\Lambda$CDM model, for which we
assume a present matter density parameter $\Omega_{0m}=0.3$, a Hubble constant
$H_0=70$ km/s/Mpc and a r.m.s. density fluctuation $\sigma_8=0.9$.  The volume
that is constrained by the observational data covers a sphere of radius $\sim
80$ Mpc/h, centered on the Milky Way. This region is sampled with more than 50
million high-resolution dark matter particles and is embedded in a periodic
box of $\sim 240$ Mpc/h on a side. The region outside the constrained volume 
is filled with nearly 7 million low-resolution dark matter particles, allowing 
a good coverage of long-range gravitational tidal forces.

The statistical analysis made by \citet{Mathis:2002} showed that, using this
technique, the simulated $z=0$ matter distribution provides a good match the
large-scale structure observed in the local universe.  Moreover, many of the
most prominent nearby galaxy clusters like Virgo, Coma, Pisces-Perseus and
Hydra-Centaurus, can be identified directly with halos in the
simulation. Therefore the galaxy clusters in this simulations are an ideal
target to study the diffuse stellar component, since observations are mainly
bound to study objects in the local universe.

Unlike in the original simulation made by \citet{Mathis:2002}, where only the
dark matter component is present, here we want to follow also the gas and
stellar component. For this reason we extended the initial conditions by
splitting the original high-resolution dark matter particles into gas and dark
matter particles having masses of $0.48 \times 10^9\; M_\odot/h$ and $3.1 \times
10^9\; M_\odot/h$, respectively; this corresponds to a cosmological baryon
fraction of 13 per cent. The total number of particles within the simulation
is then slightly more than 108 million and the most massive clusters is
resolved by almost one million particles.
The gravitational force resolution (i.e. the comoving softening length) of
the simulation has been fixed to be 7 kpc/h (Plummer-equivalent), fixed in
physical units from z=0 to z=5 and then kept fixed in comoving units at higher
redshift.

To get a handle on possible effects of resolution on our modified {\small SUBFIND}
identification scheme for the {\it cD} and the {\it DSC}, we also analyzed two
re-simulations of a massive galaxy clusters at higher resolution with exactly
the same physics included. In the high resolution run (labeled as {\it 3x} in the
following) the individual particle
masses are a factor of 3 times smaller than in our local universe simulation,
and we scaled the softening accordingly to 5 kpc/h. Besides the most massive
mail halo ($M_{vir} \approx 2\times10^{15} M_\odot/h$), this re-simulation hosts
several less massive halos ($M_{vir} \approx 1 \times 10^{14} M_\odot/h$),
which we analyzed as well. In the extreme high resolution run (labeled as {\it 18x} 
in the following) the individual particle masses are 18 times smaller and the
softening is accordingly set to 2.5 kpc/h. Here, due to additional optimizations
to the high resolution region of the re-simulation, only the main halo is suited
to be analyzed. 
This cluster is resolved by more than respectively  4 and 25 million particles within
the virial radius in the {\it 3x} and the {\it 18x} simulation. 

Our simulations have been carried out with {\small GADGET-2}
\citep{springel2005}. The code uses an entropy-conserving formulation of
SPH \citep{2002MNRAS.333..649S}, and allows a treatment of radiative
cooling, heating by a UV background, and star formation and feedback
processes. The latter is based on a sub-resolution model for the
multiphase structure of the interstellar medium
\citep{2003MNRAS.339..289S}. The code can also follow the pattern of
metal production from the past history of cosmic star formation
\citep{2004MNRAS.349L..19T,2007MNRAS.382.1050T}. This is done by computing the contributions 
from both Type-II and Type-Ia supernovae and energy feedback and
metals are released gradually in time, accordingly to the appropriate
lifetimes of the different stellar populations. This treatment also
includes, in a self-consistent way, the dependence of the gas cooling on
the local metalicity. The feedback scheme assumes a Salpeter IMF
\citep{1955ApJ...121..161S}, and its parameters have been fixed  to get 
a wind velocity of $\approx 480$ km/s.  

We note that our present local universe simulation has additional physics
with respect to M04 work. Their mass
resolution was $4.62 \times 10^9 h^{-1}$ M$_\odot$ for a Dark Matter (DM)
particle and $6.93 \times 10^8 h^{-1}$ M$_\odot$ for a gas particles. The
force resolution was $7.5 h^{-1}$ kpc (Plummer-equivalent softening length).
While force and mass resolution are similar to those of our local universe
simulation, M04 didn't include chemical enrichment and metal-dependent
cooling. 


\subsection{Substructure detection} \label{sec:detection}

Substructures within halos are usually defined as locally over-dense,
self-bound particle groups identified within a larger parent halo. In our
analysis, the identification of these substructures is performed by applying a
modified version of the {\small SUBFIND} algorithm
\citep{2009MNRAS.tmp.1314D}, which, in contrast to its original version
\citep{2001MNRAS.328..726S} works with different particle species (e.g. dark
matter, gas and star particles). Details on the algorithm can be found in
\citep{2001MNRAS.328..726S, 2009MNRAS.tmp.1314D}.  In brief, as a first step,
we employ a standard friends-of-friends (FoF) algorithm to identify the parent
halos, with a linking lenght $=0.17$ times the mean interparticle
distance. Within each FoF group we then estimate the density of each
particle species by adaptive kernel-interpolation, using the standard SPH
approach with a given number of neighboring particles. Using an excursion set
approach where a global density threshold is progressively lowered, we find
locally over-dense regions within the resulting density field, which form a
set of substructure candidates. The outer `edge' of the substructure candidate
is determined by a density contour that passes through a saddle point of the
density field; here the substructure candidate joins onto the background
structure. In a final step, all substructure candidates are subjected to a
gravitational unbinding procedure where only the self-bound part is
retained. We use all the particles initially belonging to the substructure
candidate for evaluating the gravitational potential.  For the gas particles
we take also the internal thermal energy into account in the gravitational
unbinding procedure. If the number of bound particles left is larger than a 20
particles, we register the substructure as genuine {\it sub-halo}.  We define
the stellar component of such {\it sub-halo} as a cluster member {\it galaxy}.

Although this works well for identification of satellite galaxies, all star
particles not bound to any satellite galaxy are associated with the main halo
stellar component, as the particles are always bound to the cluster
potential. Therefore, this procedure does not split such a stellar population
into the stars belonging to the central galaxy ({\it cD}) and to the stellar
diffuse component ({\it DSC}).

Previous work (M04, M07) based on {\small
  SKID} \citep{2001PhDT........21S} introduced an empirical distance scale to
limit the part of the main halo used to calculate the potential and then
distinguish between the {\it cD} and {\it DSC} by applying an unbinding
criteria to this truncated potential. In section \ref{sec:dynamics} we will
show that the main halo stellar component has two components with two distinct
dynamics. In section \ref{sec:iclunbinding} we show that this can be used to
modify the unbinding procedure to replace empirical distance scale by an
inferred one, which optimize the splitting of the two components according to
their dynamics.

To give an impression of the dynamics of the stellar component in our
simulations we modified the {\small SPLOTCH} package \citep{Dolag08}, a
Ray--tracing visualization tool for SPH simulations. Instead of mapping a
scalar value to a color table, we mapped the three velocity components directly
into the RGB colors. To obtain a logarithmic like scaling but preserve the
sign of the individual velocity components we used a mapping based on the 
inverse of the hyperbolic sin function, e.g. $<r,g,b> \propto \mathrm{
  asinh}(v_{x,y,z})$. We calculated a smoothed density field based on all star particles
using a SPH like kernel function. The resulting Ray--tracing images, are shown in
Figure~\ref{fig:stars_ray}. The upper panel shows all stars in the high
resolution sphere. Same color indicates the same 3 dimensional velocity
direction, therefore the overall color pattern nicely reveals the large
scale velocity structure within the cosmological simulation. The larger
clusters however appear mostly white meaning that the internal velocity
dispersion overcomes the ordered, large scale velocity pattern and therefore
the color saturates into white. The lower left panel shows all stars in
galaxies (including the {\it cD}) of the most massive cluster (the one with the
yellowish halo also seen close to the center in the upper panel). This
cluster has 238 satellite galaxies above our detection threshold. The different
colors of the galaxies reflect their different orbital phase, whereas the
yellow color of the {\it cD} and some other satellite galaxies reflects the
large scale velocity field as seen in the upper panel.  
As expected, besides
the mean velocity of the cluster, there is no large scale pattern visible
in the galaxy orbits. Also the individual galaxies show a uniform color indicating
that the individual velocity dispersion of the stars within the galaxies is much
smaller than their orbital motion within the galaxy cluster. The lower right panel 
shows all stars not bound to individual galaxies. The small scale structures
remaining in the {\it DSC} are either galaxies below our detection threshold,
or the outer envelopes of large cluster members, which are not
bound to the galaxies. Notice the pink, ring like structure to the
upper left, which can be associated to the group of infalling galaxies visible
at the same position with the same color in the left panel. However, most of
the {\it DSC} appears homogeneously in white and yellow colors, indicating 
a mean velocity similar to the {\it cD} and to the large scale velocity field,
but a larger velocity dispersion.


\section{Distinguishing different dynamical stellar components} \label{sec:dynamics}

\subsection{The different stellar components}

\begin{figure}
\includegraphics[width=0.49\textwidth]{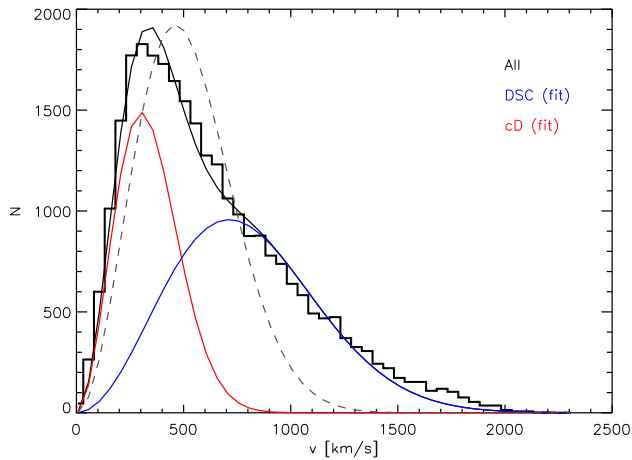}
\caption{Velocity histogram (black histogram) of the main halo stellar
component and a double Maxwellian fit to it (black line). The red and the blue
line are showing the individual Maxwellian distributions of the two components.
The gray, dashed line is our best fit with a single Maxwellian.} \label{fig:fig_1}
\end{figure}

\begin{figure}
\includegraphics[width=0.49\textwidth]{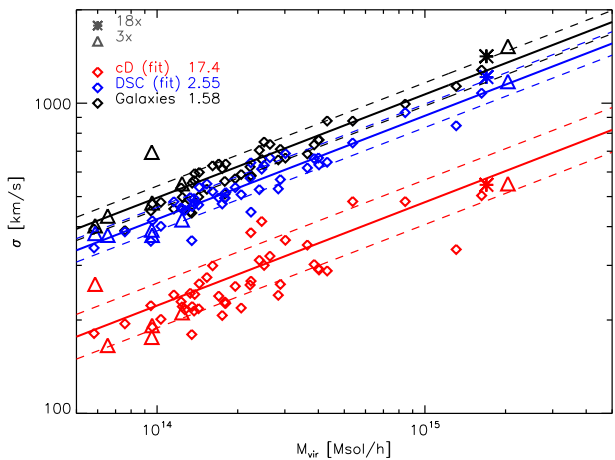}
\caption{Velocity dispersion of the three stellar components ({\it DSC}, {\it
    cD} and galaxies) as a
  function of the virial mass of the galaxy cluster for the sample of 44
  clusters with at least 20 satellite galaxies. The velocity
  dispersion of the {\it DSC} (blue symbols) is only slightly smaller than the one
  of the galaxies (black symbols), whereas the velocity dispersion of the {\it
  cD} stars (red symbols) is significantly smaller. The dashed lines show the 
  one sigma scatter of our best fit relations. Velocity dispersion shows here
  are obtained using the double Maxwellian fit \ref{eq:eq_1}. The results from the
high resolution simulations {\it 3x} and {\it 18x} are shown with triangles and 
stars, respectively.} \label{fig:fig_2}
\end{figure}

As shown in M07, the {\it DSC}  
within 0.5 $R_{vir}$ forms mainly by
merging of satellite galaxies with the {\it cD} galaxy. Therefore one could
expect that the {\it DSC} still reflects the dynamics of the satellite
galaxies and thus might have a velocity dispersion close to that of
the satellite galaxies. On the other hand, the stars belonging to the 
{\it cD} galaxy will have a much smaller velocity dispersion due to the
relaxation and merging processes, which should get rid of the excess orbital
energy in order to form the galaxy. 
Indeed, the main halo stellar component shows clear signs of a
bimodal distribution. Figure \ref{fig:fig_1} shows the histogram of the
modulus of velocities (``velocities'' hereafter) of 
the stellar component of the main halo for one of the more massive cluster 
of the simulation. This
distribution reveals its bi-modality as it can be well fitted by the 
superposition of two Maxwellian distributions
\begin{equation}
   N(v) = k_1 v^2_1 \mathrm{exp}\left(-\frac{v^2_1}{\sigma^2_1}\right) +
 k_2 v^2_2 \mathrm{exp}\left(-\frac{v^2_2}{\sigma^2_2}\right),
\label{eq:eq_1}
\end{equation}
For comparison, we also show in Figure \ref{fig:fig_1} the single Maxwellian
distribution which best fits our data. It is clear that while a single
Maxwellian is a poor fit to the velocity distribution of the stellar
component of our main halo, a double Maxwellian is a very good one.

In itself, this does not allow to identify which star particle belongs to the
{\it cD} and which to the {\it DSC} but it already allows to draw 
conclusions based on the statistical properties of the two distributions.
Therefore we selected from our
simulation all clusters having at least 20 identified satellite galaxies,
ending up with a sample of 44 galaxy clusters. For each of them we fitted a
double Maxwellian to the velocity distribution of the main halo stellar
component. Note that in all cases the fits are quite good, and introducing
a third Maxwellian component does not improve them, as the amount of star
particle associated to such third component always stays at the percent level,
and just reflects structures like the outer envelope of large, in-falling
groups.

\begin{figure}
\includegraphics[width=0.49\textwidth]{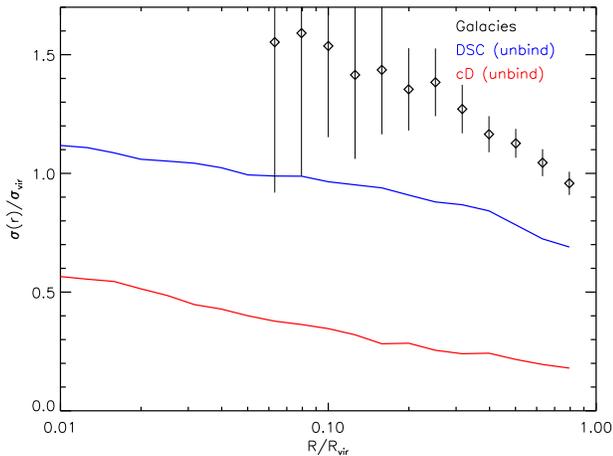}
\caption{Profiles of the radial velocity dispersions for the {\it DSC} component (blue curve),
{\it cD} component (red curve) and satellite galaxies (symbols with errorbars),
after stacking all the 44 simulated clusters of our sample. Errorbars for galaxies are
the 1$\sigma$ Poissonian uncertainties associated to the number of galaxies found
within each radial bin.} \label{fig:sigma_prof}
\end{figure}

We hypothesize that the two Maxwellian distribution of our fits correspond to
two distinct stellar components, namely the {\it cD}, associated with the
smaller velocity dispersion, and the {\it DSC} having a larger one.
Figure \ref{fig:fig_2} shows the velocity dispersion of 
such two stellar components, together with the velocity dispersion of the
galaxy population of each cluster, as a function of the virial mass of the galaxy 
cluster for our sample. The velocity dispersion of the {\it DSC} is only 
slightly smaller than the one of the galaxies: another hint on the origin of the {\it
 DSC} component from merging satellite galaxies. The slightly smaller velocity
dispersion could indicate that either some momentum transfer occurred during the merger 
event with the {\it cD}, or the galaxies which contributed to the build up of 
the {\it DSC} till redshift zero have a slightly lower velocity dispersion than
the average galaxy population. The latter could be the case, since such galaxies
might be mostly on radial orbits which lead to an earlier destruction by interactions 
with the {\it cD} than the (still) remaining galaxies within the cluster. 
However, since the {\it DSC} is more centrally concentrated than
galaxies, they could sample different region of clusters and thus have
slightly different dynamics.  
Figure \ref{fig:sigma_prof} shows the
radial velocity dispersion profile stacking all 44 clusters of our sample.
From this analysis it appears that the dynamical differences are evidently 
the same at all radii. This demonstrates that such differences are not due 
to the fact that different components sample different cluster regions.

If the creation of the {\it DSC} would be a continuous process, one would expect
a small scatter in the relation between virial mass of the cluster and velocity 
dispersions for the {\it DSC} component, specially compared to that for galaxies,
as they are much less numerous. However, from Figure \ref{fig:fig_2}, this does not 
seem to be the case, where we find a scatter of $\approx$9\% for the galaxies,
 $\approx$9\% for {\it DSC} and $\approx$18\% for the {\it cD},
 again indicating that the production of the {\it DSC}
originates in individual, violent events such as mergers with the {\it cD}. 
The velocity dispersion of the {\it cD} component is significantly lower, and
shows an even larger scatter. In Figure \ref{fig:fig_2}, solid lines represent the best fit to
\begin{equation}
   M = A \left(\frac{\sqrt{3}\sigma}{10^3\mathrm{km/s}}\right)^3\times 10^{14}M_\odot/h
\label{eq:FitM}
\end{equation}
with a normalization $A$ of 1.58, 2.55 and 17.4 for galaxies, 
{\it DSC} stars and {\it cD} stars respectively. The value of $A$ we obtain for  
galaxies is in line with the results presented in \citet{2006A&A...456...23B}. 

\subsection{The different dynamical tracers}

\begin{figure}
\includegraphics[width=0.49\textwidth]{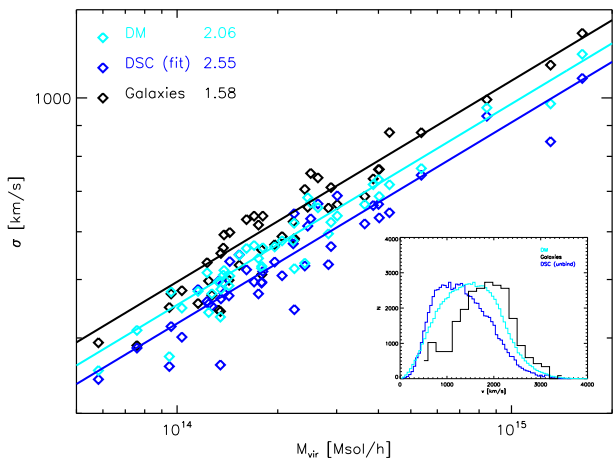}
\caption{Same as Figure \ref{fig:fig_2}, but comparing the velocity
dispersions of {\it DSC} component, galaxies and dark matter. We also report 
the value of the normalization $A$ for the best fit to equation \ref{eq:FitM}.
The inlay shows the individual velocity histograms for the three components
obtained from the most massive cluster. } \label{fig:sigma_dm}
\end{figure}

Although the differences in the velocity dispersion between the galaxies and
the {\it DSC} component as seen in Figure \ref{fig:fig_2} are small --
basically of the size of the RMS scatter of the correlation -- they seem to be
systematic.  Therefore, the question rises of which of the two distribution
traces the underlying dark matter best. Figure \ref{fig:sigma_dm} shows the
velocity dispersion of dark matter, galaxies and the {\it DSC} component for
the full set of analyzed galaxy clusters. Interestingly, the velocity
dispersion of the dark matter falls between the one of the {\it DSC} component
and the one of the galaxies. The best fit to equation \ref{eq:FitM} gives a
normalization $A$ of 1.58, 2.06 and 2.55 for galaxies, dark matter and {\it
  DSC} respectively. The inlay shows the velocity histogram (arbitrarily
normalized) of the {\it DSC}, the galaxies and the dark matter within the most
massive cluster in the simulation, clearly revealing the systematic
differences among the three different components. Note that we had to apply
our identification scheme, described in section \ref{sec:iclunbinding}, to
obtain the velocity distribution of the individual stars associated with the
{\it DSC}.

\subsection{The Mass in the different stellar distributions}

\begin{figure}
\includegraphics[width=0.49\textwidth]{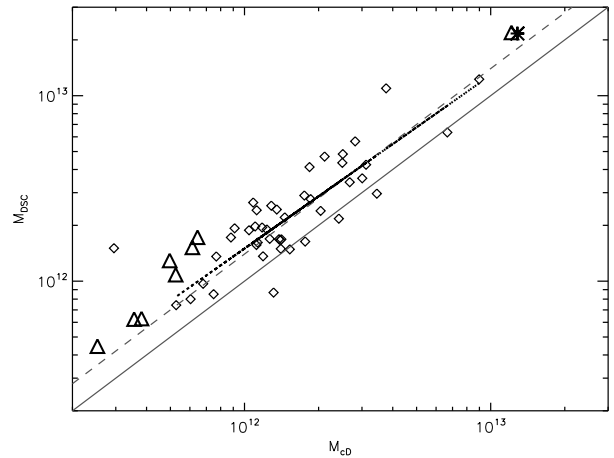}
\caption{Relation between the {\it DSC} stellar mass 
  and {\it cD} mass. The solid line marks a one to one
  relation. On average, the {\it DSC} has ca. 15\% more mass
  than the {\it cD}, as indicated by the dashed line.  
The dotted line is the best-fit linear relation, after excluding the outlier  
corresponding to the {\it cD} with mass $\simeq 3\times 10^{12}M_\odot$.
} \label{fig:fig_3}
\end{figure}

Having the velocity distribution given by our double Maxwellian fit, we can also 
evaluate the total stellar mass associated to the two individual populations by just 
integrating separately over the two distribution functions which sums up in
the double Maxwellian fit. Figure \ref{fig:fig_3} 
shows the stellar mass associated to the {\it DSC} compared to the {\it cD} 
component. From this estimate follows that the simulations predict the mass 
within the {\it DSC} to be, on average, slightly larger than the mass in the {\it cD} galaxy, 
by $\approx 15$\%, as indicated by the dashed line in figure \ref{fig:fig_3}. 
From this result follows that the mass fraction of the two dynamically identified components 
associated with the {\it DSC} and the {\it cD} will not depend on the mass of the 
galaxy cluster, being their ratio constant. 

This is different from previous findings (e.g. M04, \cite{SommerLarsen05}), and it is a feature of the 
dynamically based separation of these two components, as we will discuss in detail
later.


\section{Splitting cD and DSC components} \label{sec:iclunbinding}

\begin{figure}
\includegraphics[width=0.49\textwidth]{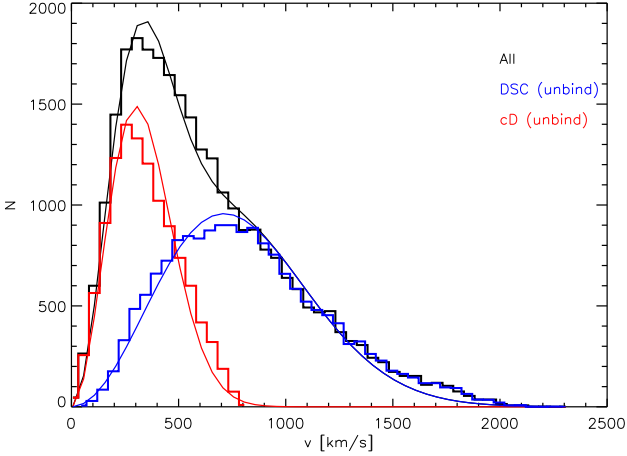}
\caption{As in figure \ref{fig:fig_1}, velocity histogram (black line) of the main halo stellar
component and a double Maxwellian fit to it (gray line) is shown. 
Red and blue histograms show the velocity distribution of the {\it cD} and
{\it DSC} stars after the un-binding procedure, together with the 
two individual Maxwellian distributions from the global fit, respectively.} \label{fig:fig_4}
\end{figure}

\begin{figure}
\includegraphics[width=0.49\textwidth]{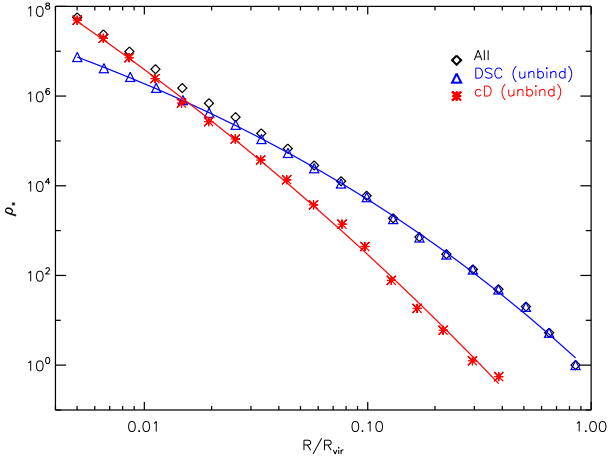}
\caption{Main halo 
  (black symbols), {\it cD} (red
symbols) and {\it DSC} (blue symbols) stellar density profiles. The red and blue line are a Sersic fit to
the two distributions, respectively.
} \label{fig:fig_5}
\end{figure}

\begin{figure}
\includegraphics[width=0.49\textwidth]{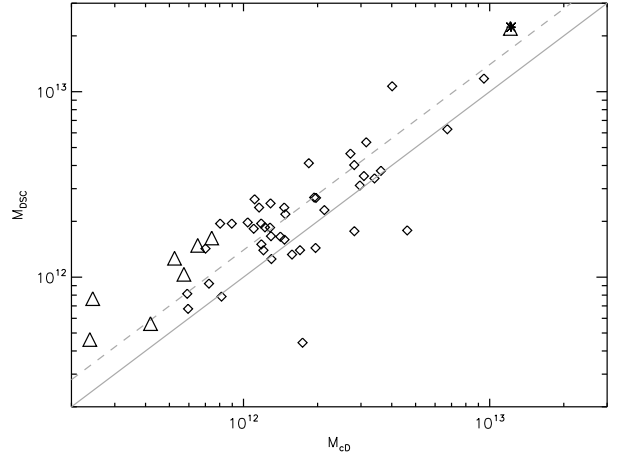}
\caption{Relation between {\it DSC}
  and {\it cD} stellar mass, after the
  unbinding. The mean relation is the same, but the
  scatter increases significantly.} \label{fig:fig_7}
\end{figure}

After detecting two distinct components, which we identify as {\it cD} and
{\it DSC}, in the velocity distribution of the stellar population, we
need a way to assign the individual star particles to one of them.

To perform this task, we start 
from a given fiducial radius (initially assumed to be a fraction
of the virial radius) and, as a first step, we calculate the gravitational potential given by all 
particles within this fiducial radius and apply the un-binding to the particles 
of the main halo component. 
As a second step, we separately fit the velocity distribution of each
of the two stellar populations, classified as ``bound'' and ``unbound'' in
the first step, to a Maxwellian distribution, obtaining two velocity
dispersions $\sigma^2_b$, $\sigma^2_{ub}$.

We compare $\sigma^2_b$ and $\sigma^2_{ub}$ to the dispersions obtained from
the original, double Maxwellian fit performed on the total star component of
the main halo, $\sigma^2_1$ and $\sigma^2_2$.  We then increase or decreases the
radius  
for the unbinding procedure, trying to match $\sigma^2_1$ and $\sigma^2_2$.  
Using this new radius and
re-computing the gravitational potential, the procedure is repeated until the
velocity distributions of the two components converges onto the the two
velocity distributions inferred from the original, global fit: $\sigma_{ub}
\cong \sigma_1$ and $\sigma_b \cong \sigma_2$.  We usually adjust the radius
to match the value $\sigma_1$ of the {\it DSC} component, but if the
algorithm does not converge, we try a second time using the dispersion
$\sigma_2$ of the {\it cD} component. The iterative procedure stops when
$\sigma_{ub}/\sigma_1$ (or $\sigma_b/\sigma_2$) differs from one less than a
given threshold value. Note that our iterative procedure tries
to match {\it one} velocity dispersion at a time; the fact that, doing so, we
also obtain the correct value of the other velocity dispersion is not given
a-priori.  Usually both dispersions agree within percent level within less
than ten iterations.

Figure \ref{fig:fig_4} shows, similarly to figure \ref{fig:fig_1}, the
velocity histogram of the main halo stellar component (black line) in the same,
massive cluster of the simulation, fitted by the superposition of two
Maxwellian distributions (solid lines).  We also show the histograms of the
bound stellar particles belonging to the {\it cD} component (red histogram) as
well as the unbound stellar component associated to the {\it DSC} (blue
histogram). Clearly, the two components are well identified and separated by
this procedure.

\begin{table}
\begin{tabular}{|l|c|c|c|c|c|}
\hline 
Component  & $\sigma$ & $I_0$  & $A$ & $\alpha$ \\ 
\hline 
{\it cD}  & 226 & 23.3 & 26.0 & 10.5 \\
{\it DSC} & 516 & 11.5 & 11.7 & 5.7 \\
\end{tabular}
\caption{
Parameters of the Sersic fit for the {\it DSC} and {\it cD} components 
(columns 1-2) and of the fit to the density profiles for the two components 
obtained from double-Maxwellian fit (see eq. 3).}
\label{tab}
\end{table}

Once  we identify the individual stars in the two distributions, we can plot their
radial density profile. Figure \ref{fig:fig_5} shows the radial density 
profiles of the {\it cD} and {\it DSC} component respectively, as well as
a fitted Sersic profile
\begin{equation}
 {\rm log}(\rho_*) = I_0-A\left(\frac{r}{R_{vir}}\right)^{(1/\alpha)}.
\end{equation}
We find that both profiles are fitted 
quite well by individual Sersic profiles and therefore the total, radial
stellar density profile can be described as a superposition of two Sersic
profiles. Also, in line with previous finding of M04, the profile of the {\it cD}
component is much steeper and less curved than that of the
{\it DSC} component. Therefore, also using our new {\small SUBFIND} identification
scheme for the {\it DSC}, we find excess light at large radii with respect to
the {\it cD} Sersic profile.  
We report in Table  \ref{tab} the values of the best-fit parameters.

Comparing the velocity dispersions $\sigma_1$,$\sigma_2$ inferred from the
original, double Maxwellian fit to all the main halo stars and the two
dispersions $\sigma_b$, $\sigma_{ub}$ for {\it cD} as well as the {\it DSC} we
find a very tight relation. In almost all cases the iteration was successfully
based on the {\it DSC} component, which led to a scatter below our chosen
threshold. Although the dispersions of the inferred {\it cD} component match
very well the corresponding ones of the double Maxwellian fit, the scatter is
slightly larger.
It is difficult to say if an improvement of
the iteration procedure could further reduce this scatter or if it is intrinsically
due to our un-binding scheme. In effect, the very fact that unbinding star particles
using the matter contained in a sphere of a given radius divides them into two
sets, whose velocity distributions match the two component of the overall double
Maxwellian distribution is not a-priori guaranteed. We regard to this match as a
hint that the double Maxwellian shape of the overall star particle
distribution is really produced by the interplay of two well separated gravitational
potentials, the cluster one and the central galaxy one.

Figure \ref{fig:fig_7} shows the relation between the stellar mass within the {\it DSC}
and the {\it cD} component,  like Figure \ref{fig:fig_3}. This time we
directly use the mass of our two components, as obtained by our unbinding scheme.
As before, the dashed line indicates an 15\% higher mass in the {\it DSC}
than in the {\it cD} component.

A larger scatter in the masses of the {\it cD} obtained from the unbinding
compared to the ones calculated from the velocity distributions is clearly
visible. Beside this, there is no bias associated with our unbinding scheme:
it seems able to disentangle two component whose properties are identical to
those inferred from the velocity distribution of the whole stellar population.

In Figure \ref{fig:fig_3},\ref{fig:fig_4} and \ref{fig:fig_7} we also report
results obtained from two individual galaxy cluster re-simulations (using
different symbols, as indicated in the plots), which have 3 and 18 times
better mass resolution. Such results are perfectly in line with those obtained
from the cosmological simulations.  Thus, our unbinding scheme appears to be
stable against resolution effects. Note that, at variance with other schemes,
e.g {\small SKID}, we don't have any parameter to be tuned when resolution is changed.
However, we excluded these points from the fits presented in Figure \ref{fig:fig_5}.


\section{Properties of the {\it DSC} in the simulated local universe} \label{sec:cDDSC}

\begin{figure}
\includegraphics[width=0.49\textwidth]{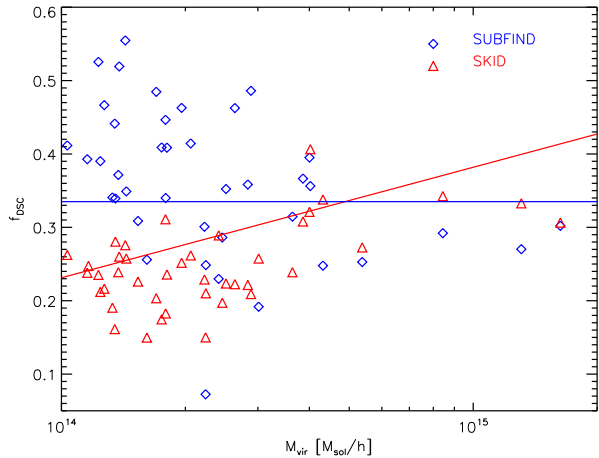}
\caption{Fraction of stellar mass in {\it DSC} as a function of the galaxy
  cluster virial mass. Blue diamonds show the {\it DSC} fractions as given by our new
  {\small SUBFIND}+un-binding algorithm, red triangles show the {\it DSC} fractions as
  obtained using {\small SKID}. The red line shows the result of a least-chi-squared
  fit of the {\small SKID} fractions to the linear function $F_{DSC} = b \cdot M_{clus} + a$, which gives
  $a=0.211$, $b=1.03 10^{-16}$, $\sigma_a=0.01$, $\sigma_b=2.59 10^{-17}$,
  $\chi^2 = 0.112$. Performing the same fit using the {\small SUBFIND} fractions gives a
  value of $b$ compatible with zero at 1.5 $\sigma$, we thus fit {\small SUBFIND} {\it DSC}
  fractions to a constant function $b=0.335 \pm 0.02$ (blue line) 
} \label{fig:fig_8.5}
\end{figure}

\begin{figure}
\includegraphics[width=0.49\textwidth]{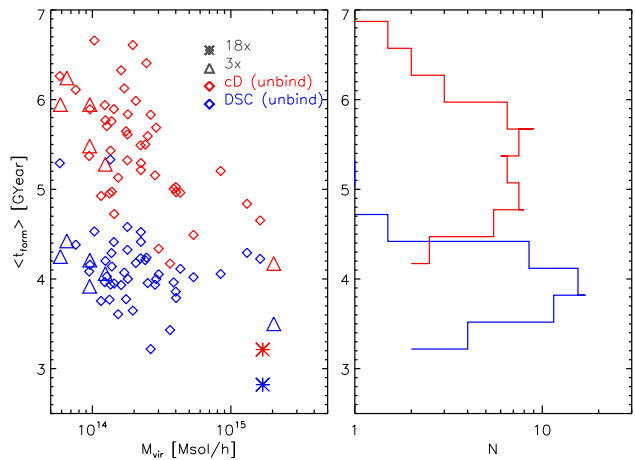}
\caption{The left panel shows the mean formation times of {\it cD} (red symbols) and {\it
    DSC} (blue symbols) as function of the virial mass of the cluster. Results from the higher 
resolution, {\it 3x} and {\it 18x}, simulations, are shown with triangles and stars, respectively. The
  right panel shows the histograms of the distribution of the mean formation
  times for the {\it cD} and the {\it DSC} components, respectively.} \label{fig:fig_10}
\end{figure}

\begin{figure}
\includegraphics[width=0.49\textwidth]{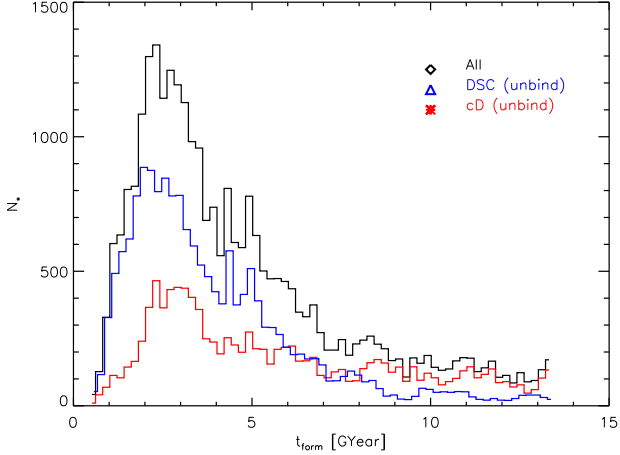}
\caption{Formation time of main
  sub-halo stars (black histogram), {\it cD} stars (red histogram) and
  {\it DSC} stars(blue histogram) in the most massive cluster in the
  cosmological simulation.
} \label{fig:fig_9}
\end{figure}

We now apply our new un-binding scheme to all 44 massive clusters in the local
universe simulation with at least 20 member galaxies. In all cases, Sersic
profiles are a good fit to the radial density profiles of {\it cD} and {\it DSC}
populations separately, as already shown for one massive cluster in Figure
\ref{fig:fig_5}. Also, as expected, the {\it cD} component in all systems is
radially more concentrated than the {\it DSC}.  
By inspecting the radial 
density profiles of the separated components 
we find that the {\it DSC} component in all
cases dominates at radii larger than 0.03-0.08 $R_{vir}$ and contributes less
than 10\% to the stellar density in the center.

\subsection{Comparison with previous works} \label{sec:bah}

Figure \ref{fig:fig_8.5} shows the fraction of stellar mass in the {\it DSC}
against the cluster virial mass. To make a significant comparison with M04
results, we also analyzed the local universe simulation using {\small SKID}, repeating
the same procedure described there (see also Appendix for details). We remind
the reader that M04 analyzed a cosmological box of size $192 h^{-1}$ Mpc, run with mass
and force resolution similar to those used here, but different physics: in
particular, they didn't follow chemical enrichment and didn't use a
metal-dependent cooling function.

Nevertheless, we recover the trend of having a larger {\it DSC} fraction in
larger clusters when we define it using {\small SKID}, even if such a trend is weaker
here than in M04.  Our new dynamical identification of the {\it DSC}, however, 
gives a different results. Although the {\it DSC} fractions for the most massive cluster 
agree well, on average they are higher than those found using {\small SKID},
especially at low cluster masses, and, moreover, we can't detect any trend with the
cluster mass. In Appendix, we show a detailed comparison of one high mass and one
low mass cluster analyzed with both methods to demonstrate that this effect is indeed
related to the fact, that for the {\small SKID} analysis a scale has to be given
a-priori, which could lead to this effect. However, we want to stress once more 
that the {\it DSC} mass fraction found here should not be regarded as
immediately comparable to observations. Especially in low-mass clusters, the
{\it DSC} abundance is expected to be dominated by central regions. Central
regions are exactly those where a dynamical identification of a {\it DSC}, as
that presented here, can most differ from a surface brightness based detection
of an {\it ICL} components. Dynamical analysis of the {\it DSC} should instead
be regarded as attempts to determine its physical properties and its origin.

\subsection{Different stellar populations in cD and DSC} \label{sec:pop}

In line with M04, we find that the stellar populations of the two different
components, however we identify them, differ significantly not only in their
spatial distribution, but also in their history.

The left panel of Figure \ref{fig:fig_10} shows mean formation times of the stars
in the {\it cD} (red symbols) and {\it DSC} (blue symbols) for our sample of
44 clusters as obtained using our new procedure. In the {\it DSC}  
(and therefore 
its progenitor galaxies), there is
no significant star formation within the last 4-5 Gyrs, while the {\it cD}
population is on average younger. The difference in the two age distributions
is also clear from the right panel of Figure \ref{fig:fig_10}, where we show
the histograms of mean formation times for the {\it cD} (red) and {\it DSC}
(blue). This agrees with the finding of M07,
that {\it DSC} is mainly formed during merger processes which also form the
central galaxy of each cluster. Destroyed satellite galaxies, and
pre-processed stars, already unbound in merging galaxy groups, goes to the
{\it DSC} of clusters, whereas the {\it cD} still can form stars at low rate
at recent times. In numerical simulation of galaxy clusters, one
known issue is the excess star formation at the center of cooling flows,
producing an overly blue central galaxy, see e.g \cite{2006MNRAS.373..397S}. 
Real BCGs have a much lower low-z star formation rate, therefore we
expect this age segregation to be smaller in reality; by how much, is stil
to be completely understod.
Note, however, that this mechanism applies to {\it all} mergers at {\it all}
times. Therefore, also at higher redshifts, the merger remnants at the center
of proto-clusters will have
ongoing star formation while, obviously, the mean age of the produced {\it DSC} will
grow older. This can be seen from Figure \ref{fig:fig_9} 
where we show, for the most massive cluster in the
cosmological simulation, the distribution of formation times of star particles
for all stars in the main sub-halo, {\it cD} stars and {\it DSC} stars. Star
particles older than 7 Gyr (i.e., $t_{\rm form}<6$) are abundant in the {\it DSC}; these stars formed
at redshifts $z>1$ and must have become unbound before. At late-times,  
lack of feedback in the simulations leads to ongoing star formation at the center of the cluster 
gives an excess of young stars ($t_{\rm form}>8$ Gyr) in the {\it cD}. 
However, even ignoring
all stars with $t_{\rm form}>8$ Gyr, the difference between the average formation time 
of {\it cD} and {\it DSC} still remains half of the effect shown in 
figure \ref{fig:fig_10}. 

As before, we added in figure \ref{fig:fig_10} the results obtained using our
two re-simulations of individual galaxy clusters.  Again, they reproduce the
trends found in our cosmological simulations, and for the same reasons as
before, we did not include them in the histograms.  Note that, especially in
the very high resolution run, star formation is slightly shifted to earlier
times, as expected since such a simulation resolves the first star-forming
structures better.

We also evaluated mean formation times of {\it cD} and {\it DSC}
stellar populations given by {\small SKID} for our 44 clusters. The
result is similar to that shown for {\small SUBFIND} in Figure
\ref{fig:fig_9}, but mean formation times of the two components are
closer with the average formation time of {\it cD}s being 4.9 Gyr and
that of {\it DSC}, 4.1 Gyr. Again, this suggests that {\small SKID} is
less efficient in disentangling the two components, and assign to some
{\it cD} star particles which have properties typical of the {\it
 DSC}.

\subsection{Origin of the scatter of the DSC} \label{sec:origin}

\begin{figure}
\includegraphics[width=0.49\textwidth]{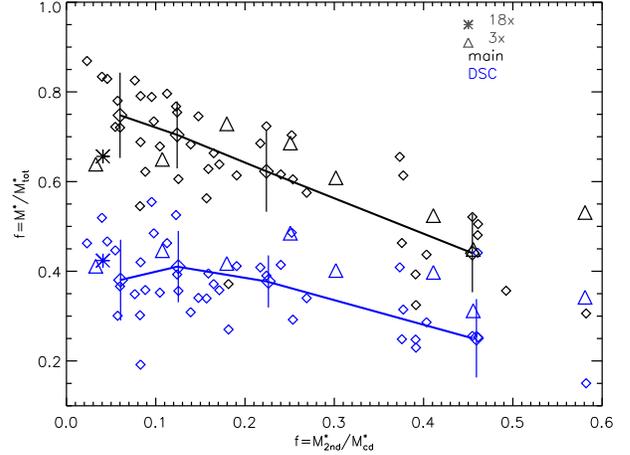}
\caption{Fraction of stellar in  the 
main sub-halo, the {\it DSC} and the {\it cD} component, respectively, to the
total mass in stars as a function of the 
fraction of the stellar mass of the second brightest galaxy in each cluster to
the {\it cD} mass. The results from the
high resolution simulations {\it 3x} and {\it 18x} are shown with triangles and 
stars, respectively.} \label{fig:fig_12}
\end{figure}

\begin{figure}
\includegraphics[width=0.49\textwidth]{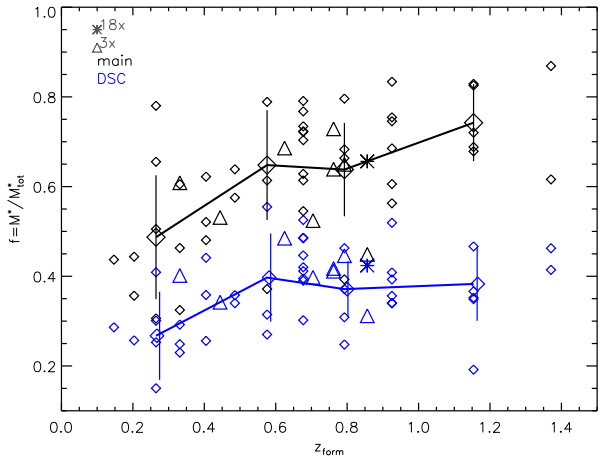}
\caption{Same as figure \ref{fig:fig_12} but as a function of the
formation redshift, defined as the redshift at which the mass of the
main progenitor is half of the mass of the descendant cluster at redshift zero.
The results from the
high resolution simulations {\it 3x} and {\it 18x} are shown with triangles and 
stars, respectively.} \label{fig:fig_13}
\end{figure}

Both observations and previous numerical works indicated that the fraction of
the {\it DSC} component in clusters having similar mass shows a large scatter.
Reminding the reader that our cluster set is a complete, e.g. volume limited sample, we
can try to investigate where the scatter in relations of {\it cD} and {\it
 DSC} components with the cluster mass comes from. Thus,  we calculated the 
the total mass for our different stellar components, namely the total amount of 
formed stars, the amount of stars in the main sub-halo and the stellar mass of 
{\it DSC} and {\it cD} components respectively. 
In line with previous findings \citep{2006MNRAS.367.1641B},
there is a week trend to form more stars in smaller clusters, but the scatter of formed 
stars in similar mass systems is quite small. Although the global trends is reflected
more or less homogeneously in all different components, the relative scatter, especially 
in the main and {\it DSC} component is quite substantial.

As seen before, the un-binding procedure induce additional scatter by itself;
however this does not account for the whole variation of the {\it DSC}
fraction among different systems of similar mass. Figure \ref{fig:fig_12}
shows the fraction of all stars in the main sub-halos and in the {\it DSC}, as
a function of the relative importance of the second most massive galaxies
within each cluster.  There is a clear indication that the presence of a quite
massive second galaxy (indicating a young system, where the remain of the last
major merger has not yet merged with the {\it cD}) is related to a less
prominent {\it DSC} component.  
Note that this is fully compatible with the
low fraction of {\it ICL} observed in the Virgo cluster ($<10$\%), 
which is characterized by the presence of two comparable 
brightest cluster galaxies. Therefore
the observed Virgo cluster lies even to the right of the shown 
range in figure \ref{fig:fig_12}. 

By constructing the merger tree of our set of clusters we can trace back the main
progenitor of each of the cluster. Adopting the common definition for the
redshift of formation of a cluster $z_\mathrm{form}$  (redshift at which 
the main progenitor of the cluster has half of its final mass) we can check
how the stellar fractions in the different components depend on the age of the 
cluster. Figure \ref{fig:fig_13} shows the stellar fraction of the main sub-halos and 
the {\it DSC} component as a function of the formation redshift $z_\mathrm{form}$.
In old systems, expected to be very relaxed, the growth of the {\it cD} galaxy seems to
be the dominant source of increasing the stellar fraction, whereas the {\it DSC} 
seems not be evolving significantly. Only in very young systems the {\it DSC}
shows a deficit, most likely because the merging is still ongoing and the 
{\it DSC} is not yet released from the progenitors of the forming systems.

Summarizing, when we apply our new un-binding scheme, a correlation of the
{\it DSC} with the cluster mass is not present at all. The amount of diffuse
stars appears to depend more on the formation redshift of the cluster and on
its dynamical state: more relaxed  
and evolved clusters have more {\it DSC} because they
had more time to complete the merging of galaxies onto the {\it cD}.


\section{Conclusions} \label{sec:conc}

We showed that the stellar populations of the main sub-halos of massive galaxy
clusters in cosmological, hydrodynamical simulations are composed of two,
dynamically clearly distinct components. Their velocity distributions can be
fitted by a double Maxwellian distribution. Including this information into an
un-binding procedure, we where able to spatially separate the two components
in a central galaxy ({\it cD}) and a diffuse stellar component ({\it
  DSC}). The latter can be associated with the observationally very
interesting intra cluster light ({\it ICL}) component.
However, we want to stress that there are significant differences in how the {\it ICL}
is measured in observations and how the {\it DSC} is inferred in simulations. Therefore
a detailed comparison of the amount of this stellar component of galaxy clusters
in simulations and observations can only be done by mimicking observational strategies, 
e.g. inferring the component from synthetic surface brightness maps. 

On the other hand
our un-binding algorithm is ideally suited to study the physical processes leading to 
liberate the {\it DSC} as it allows to distinguish the {\it DSC} and the {\it cD}
with a minimum set of assumptions, namely
\begin{itemize}
\item The stars of the {\it cD} component are those gravitationally bound to the inner 
      most part of the halo, i.e. to the galaxy itself.
\item The stars of the {\it cD} and the {\it DSC} reflect two dynamical distinct
      populations.
\end{itemize}
These two assumptions naturally reflect our expectations that the {\it cD}
galaxy is a relaxed object sitting in the center of the cluster potential, while
{\it DSC} is mainly formed by violent mergers of satellite galaxies with the 
{\it cD}, is in equilibrium with the overall gravitational potential of the
cluster, and still holds memory of the dynamics of the satellite galaxies.

We show that our un-binding scheme is able to disentangle two stellar
populations, {\it cD} and {\it DSC}, whose velocity distributions
separately matches the two components of the double Maxwellian velocity distribution
which characterizes the stars belonging to the main sub-halos of clusters.

Applying our algorithm to a cosmological, hydrodynamical simulation of the local 
universe including cooling and star formation, we found that 
\begin{itemize}
\item The velocity dispersion associated to the {\it DSC} component is comparable
to the velocity dispersion of the member galaxies, whereas the velocity dispersion
associated to the {\it cD} is significant smaller (by a factor of $\approx
3$). his implies that the {\it DSC} dynamics is determined by the general
gravitational potential of the cluster, while the {\it BCG} one feels the
local galactic potential.

\item In line with previous findings, the spatial distribution of the star particles 
belonging to the {\it cD} are much more concentrated than the one of the {\it DSC}, 
which starts to dominate the stellar density at radii larger than 0.03-0.08 $R_{vir}$.
\item The formation time of the star particles associated with the {\it cD} is 
significant smaller than the one of the {\it DSC}, indicating that -- in line with 
previous findings -- the {\it DSC} is older (on average $\approx 1.5$Gyears).
\item The fraction of the stars within the {\it DSC} is only weekly dependent on
the mass of the galaxy cluster, however, we find a larger fraction in clusters
with a early formation time. This is in line with previous findings that the {\it DSC} 
originates from the late stages of merger events of galaxies with the {\it cD}. 
Therefore clusters which have had enough time to liberate stars since the last major 
infall show a larger {\it DSC}. Interestingly, the mass ratio between the second 
brightest cluster galaxy and the {\it cD} galaxy can be used as a proxy for the time
till the last major merger. Using such a ratio, we obtain an even stronger trend:
clusters with a small ratio show a large {\it DSC}
fraction, whereas clusters with a large ratio did not have yet enough time to 
liberate the stars contribution to the {\it DSC}. We expect that such a trend
could also be detected in {\it ICL} observations.
\end{itemize}

We conclude that the separation of the {\it cD} and the {\it DSC} in simulations, 
based on our dynamical criteria, is more opportune than other current methods,
as it depends on less numerous and physically more motivated assumptions, and 
leads to a stable algorithm helping to understand the different stellar components 
within galaxy clusters.


\section*{acknowledgments}
The hydrodynamical simulations on which the presented work is based on have been
performed using computer facilities at the University of Tokyo supported by
the Special Coordination Fund for Promoting Science and Technology, Ministry
of Education, Culture, Sport, Science and Technology.  K.~D.~acknowledges the
supported by the DFG Priority Programme 1177, the hospitality of the Department
of Astronomy of the University of Trieste, the receipt of a ``Short Visit
Grant'' from the European Science Foundation (ESF) for the activity entitled
``Computational Astrophysics and Cosmology'' and the financial support by the
``HPC-Europa Transnational Access program''. This work has been partially
supported by the PRIN-MIUR grant "The Cosmic Cycle of Baryons", by the
INFN-PD51 grant and by a ASI-COFIS theory grant.

\bibliographystyle{mn2e}
\bibliography{master,master3}

\appendix

\section{Skid -- Subfind comparison}

\begin{figure*}
\includegraphics[width=0.99\textwidth]{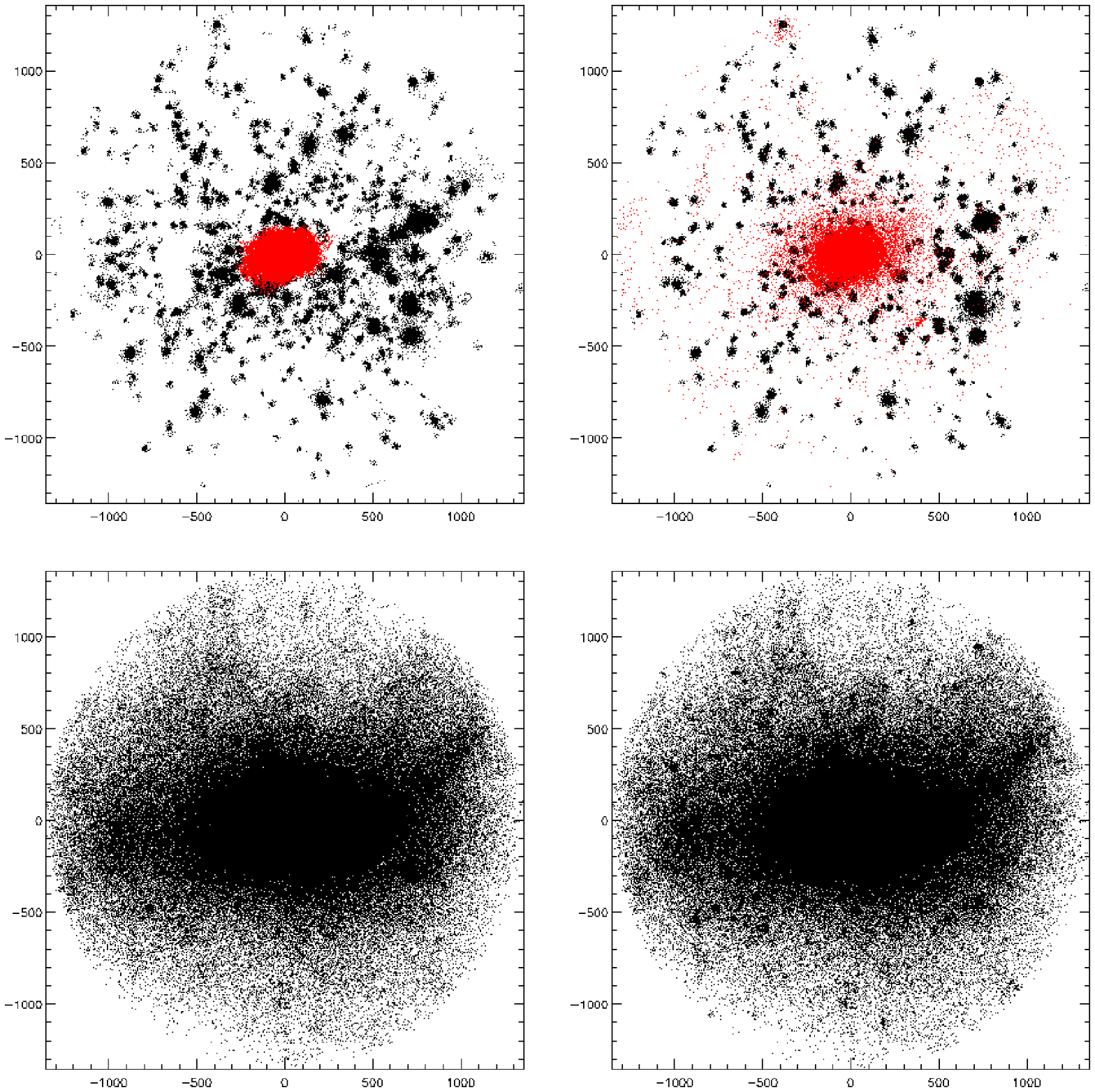}
\caption{All star particles within $R_{vir}/2$ of the high resolution ({\it 3x})cluster belonging
to individual galaxies (black dots in the upper row), the {\it cD} component (red dots in the upper row)
and the {\it DSC} component (lower row). The left column is obtained applying {\small SKID}, the right column is
our {\small SUBFIND} result.
} \label{fig:comp_1}
\end{figure*}

\begin{figure}
\includegraphics[width=0.49\textwidth]{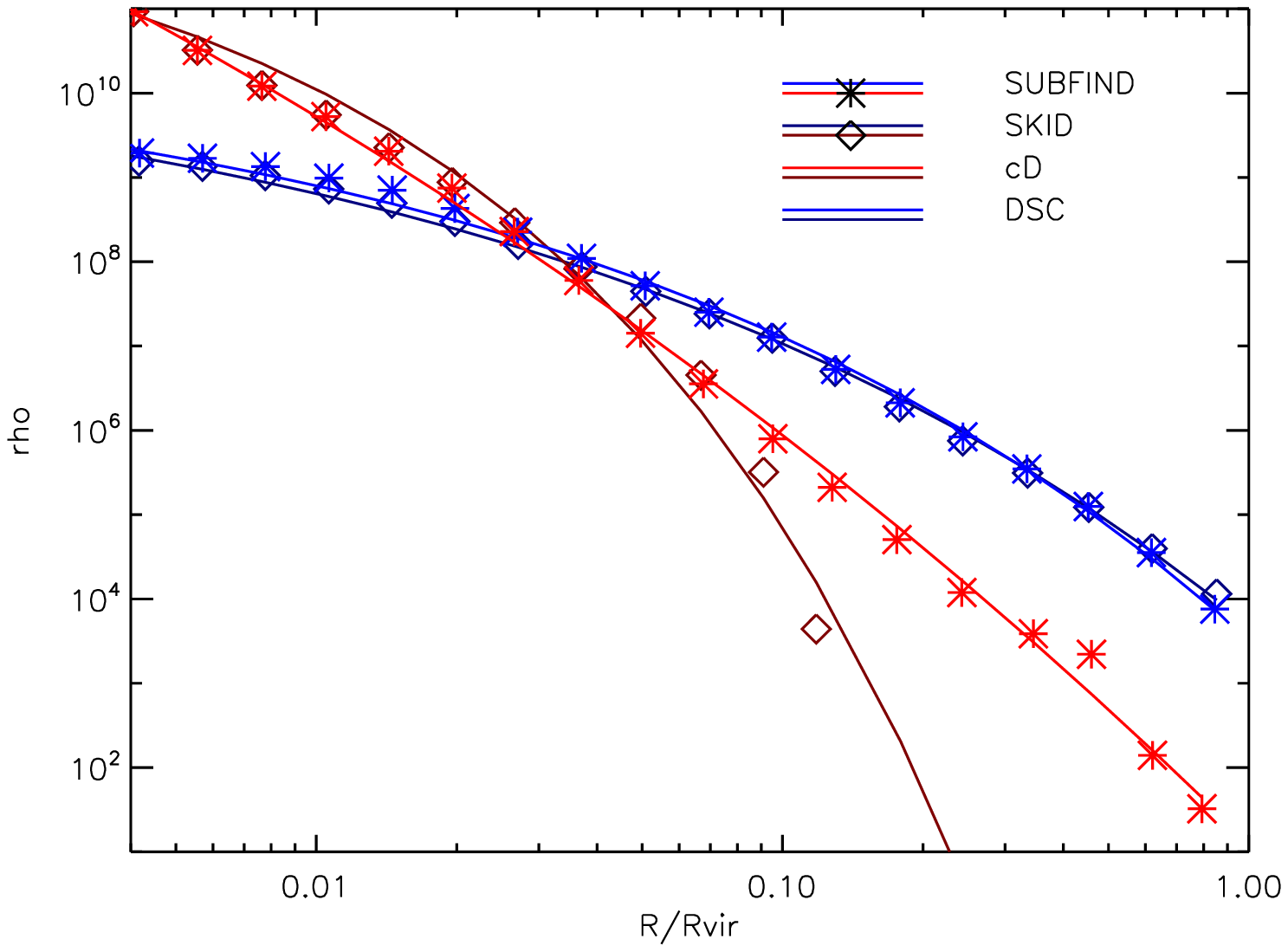}
\caption{As in Figure \ref{fig:fig_5}, the stellar density profile of the 
{\it cD} (red symbols) and the {\it DSC} (blue symbols) are shown. 
The red and blue lines represent a Sersic fit to
the distributions. Diamonds with solid lines and stars with dashed lines 
are the results obtained with {\small SKID} and {\small SUBFIND}
respectively.
} \label{fig:comp_2}
\end{figure}

\begin{figure}
\includegraphics[width=0.49\textwidth]{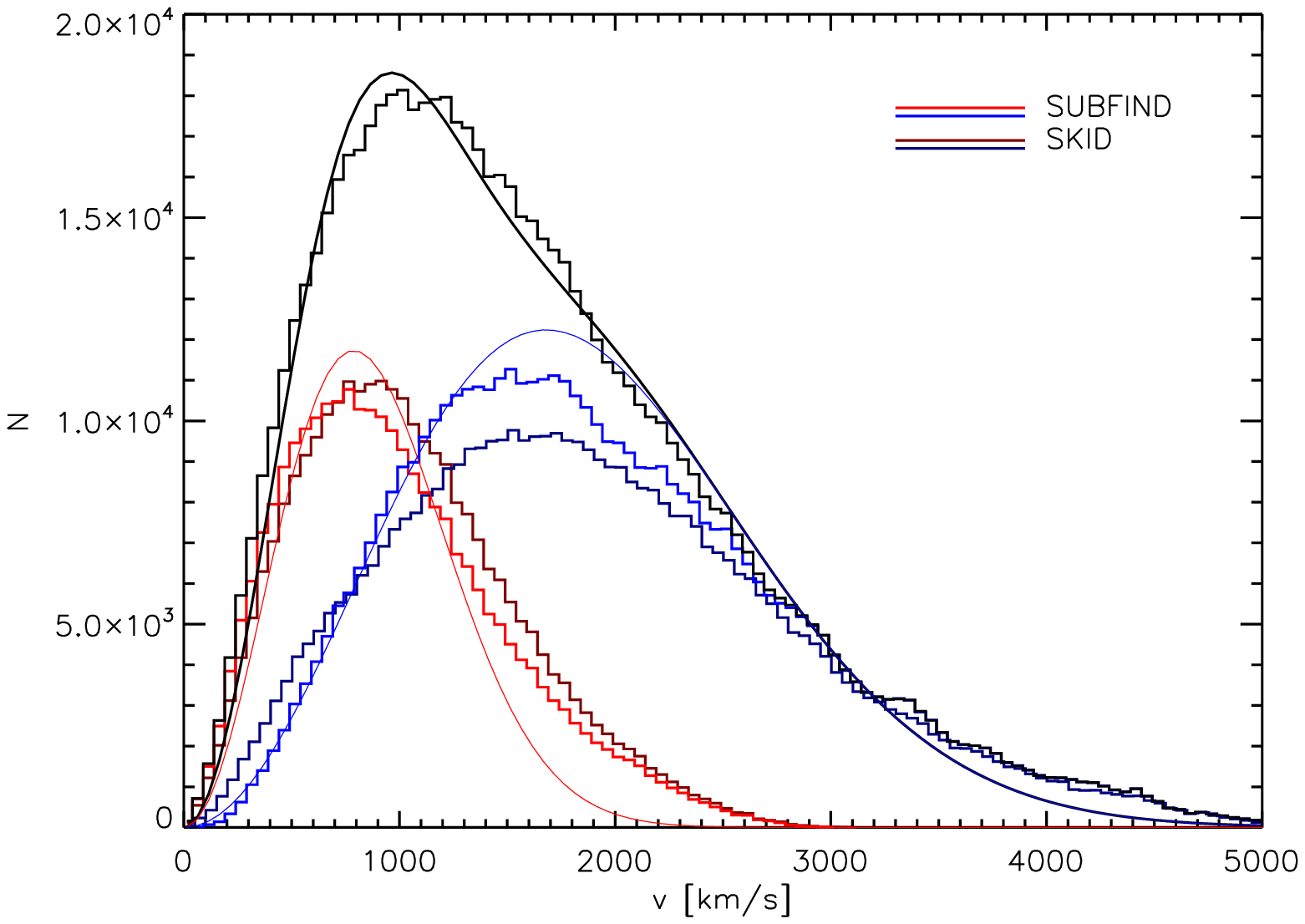}
\caption{As in Figure \ref{fig:fig_4}, velocity histogram (black histogram) of the main halo stellar
component and a double Maxwellian fit to it (thin line) is shown. Additionally,
the red and blue histograms show the velocity distribution of the {\it cD} and
{\it DSC} components and the thin lines indicate the individual, Maxwellian distributions
obtained from the global fit. Light and dark colored histograms are the results obtained with 
{\small SKID} and {\small SUBFIND}
respectively.
} \label{fig:comp_3}
\end{figure}

The identification of galaxies in 
M04 and M07 was based on a {\small SKID}.
\citep{2001PhDT........21S}. For obtaining stable results, {\small SKID}
parameters must be tuned carefully for correctly identifying individual
galaxies. Namely, M07 showed that three different smoothing length must be
used, and the results of the three analysis combined, to avoid spurious loss
of low density galaxies. Moreover, in SKID analysis, a typical
scale length $\tau$ must be supplied for performing the un-binding
procedure. Here (as in M04 and M07), we linked such a scale to the gravitational
softening $\epsilon$ of the simulation: $\tau=3 \epsilon$. M07 showed that
this choice is well suited to obtain a reasonable separation between the {\it
  cD} and the {\it DSC}.
  
The differences found between the {\small SUBFIND} and {\small SKID} for the {\it
  cD} vs. {\it DSC} separation as a function of mass, could be due to the fact
that un-binding in clusters of different masses uses the same $\tau$. In
smaller clusters, more {\it DSC} particles could be improperly assigned to
central galaxies. Such a problem is not present in our new {\small SUBFIND}
scheme.

One additional small but systematic difference is in previous studies based on
{\small SKID} where always based on all star particles within the virial
radius, whereas {\small SUBFIND} operates on all particles belonging to a FoF
group.  The linking length is usually chosen to correspond to the virial
over-density in a given cosmology and therefore the total mass or the amount
of stars is quite similar in both analysis, but the FoF group usually is
elliptically elongated and therefore the particles sets the two algorithms
start from are not exactly the same in the outer region of the cluster.

\subsection{A detailed case study for a massive cluster}

Here we compare the analysis of the high resolution simulation (e.g. the {\it
  3x} run) applying both methods to check the differences details when
applying the two algorithm.

Figure \ref{fig:comp_1} shows a direct comparison of all star particles in the 
different components as identified by {\small SKID} (left column) and 
{\small SUBFIND} (right column) within half of the virial radius of the cluster. 
It is quite encouraging that practically all galaxies have counterparts in both methods. 
In general, it seems that the un-binding applied by {\small SUBFIND} tend to unbind 
a little bit more the outer envelope of satellite galaxies, which in return appear as 
small over-densities in the {\it DSC} component. On the contrary, the {\it cD} identified
by {\small SUBFIND} has smoother boundaries, while it looks more truncated in the {\small SKID}
analysis. Nevertheless, the mass of the {\it cD} component is very similar, namely 
$1.28\times10^{13}M_\odot/h$ versus $1.21\times10^{13}M_\odot/h$ in {\small SKID} and 
{\small SUBFIND} respectively. Note that the {\it cD} velocity dispersion is
similar in both analysis. {\small SKID} truncates the galaxy simply because its
farther stars are not assigned to the {\it cD}, but to some satellite galaxy to
which they don't belong, during {\small SKID} FoF phase. Then, they become
unbound in the un-binding phase. {\small SUBFIND} assign such stars to the main
sub-halo, and lately, they are {\it not} unbind from the {\it cD}.

The {\it DSC} looks extremely similar in both algorithms, which is also
reflected in a quasi indistinguishable density profile as shown in figure
\ref{fig:comp_2}. Here also the {\it cD} component shows excellent agreement
in the inner part, again the differences in the outer parts are visible.  We
note that also in the outer parts the {\it DSC} components in both methods
fall on top of each other, despite the previously discussed small differences
in the underlying particle sets.

Figure \ref{fig:comp_3} shows the velocity distribution in the two
components. Already in the {\small SKID} analysis the difference in the
dynamics of the two components are visible.  Although again very similar, the
{\small SUBFIND} results are more closely following the decomposition of the
main stellar component into two Maxwellian distributions, as expected. Again,
despite the different underlying algorithms, the total velocity distributions
of all star particles used in {\small SKID} and {\small SUBFIND} is quasi
indistinguishable and also leads to the same double Maxwellian fit. Therefore,
we plotted in figure \ref{fig:comp_3} only the total distribution and the
Maxwellian fits to {\small SUBFIND} data to avoid confusions.

\subsection{An example of a low mass cluster}

\begin{figure*}
\includegraphics[width=0.99\textwidth]{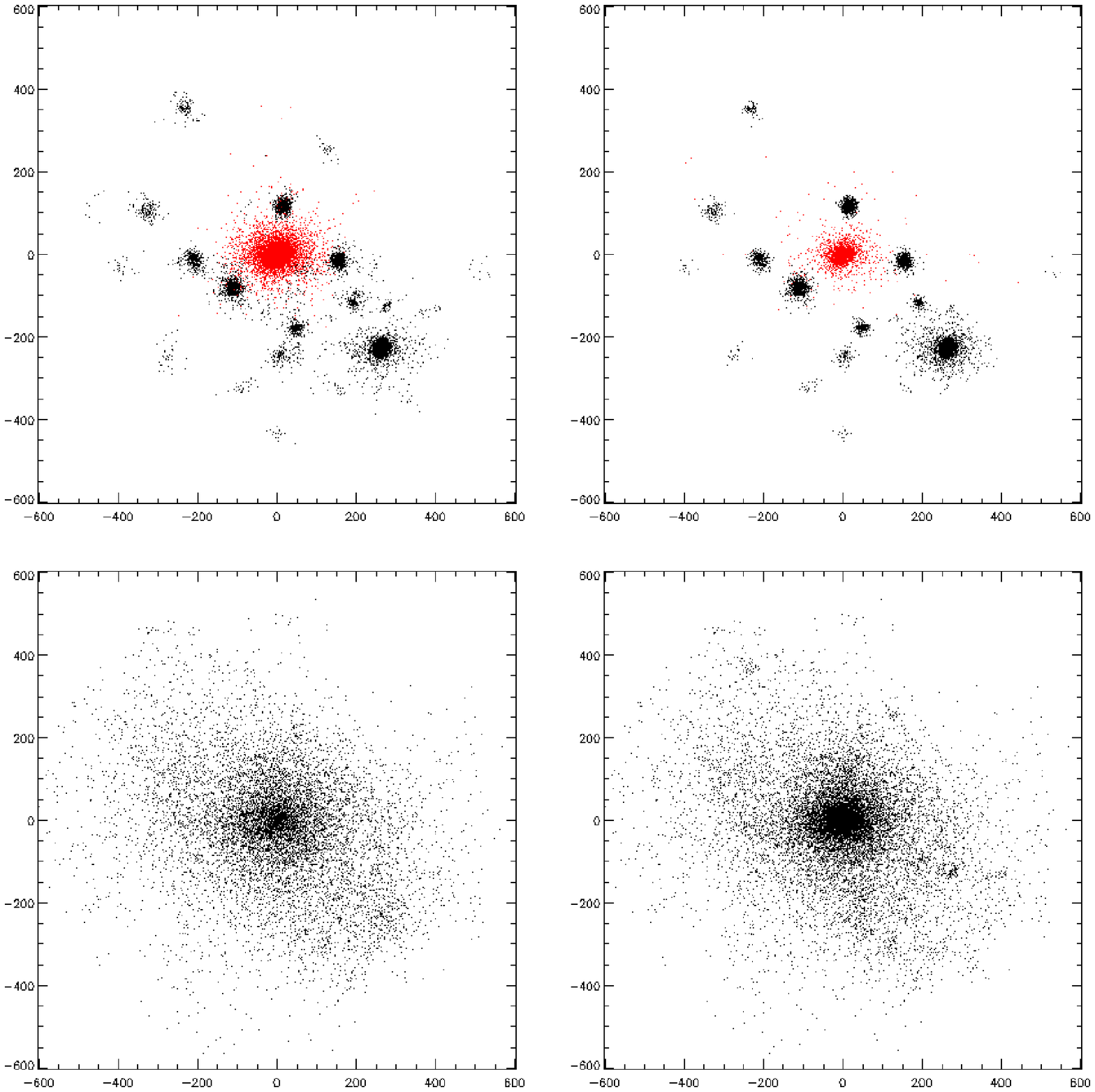}
\caption{Same as figure \ref{fig:comp_1}, but for the small cluster where we found
significant differences between the {\it cD} and {\it DSC} separation when applying
{\small SKID} (left column) and {\small SUBFIND} (right column) respectively.
} \label{fig:comp_4}
\end{figure*}

\begin{figure}
\includegraphics[width=0.49\textwidth]{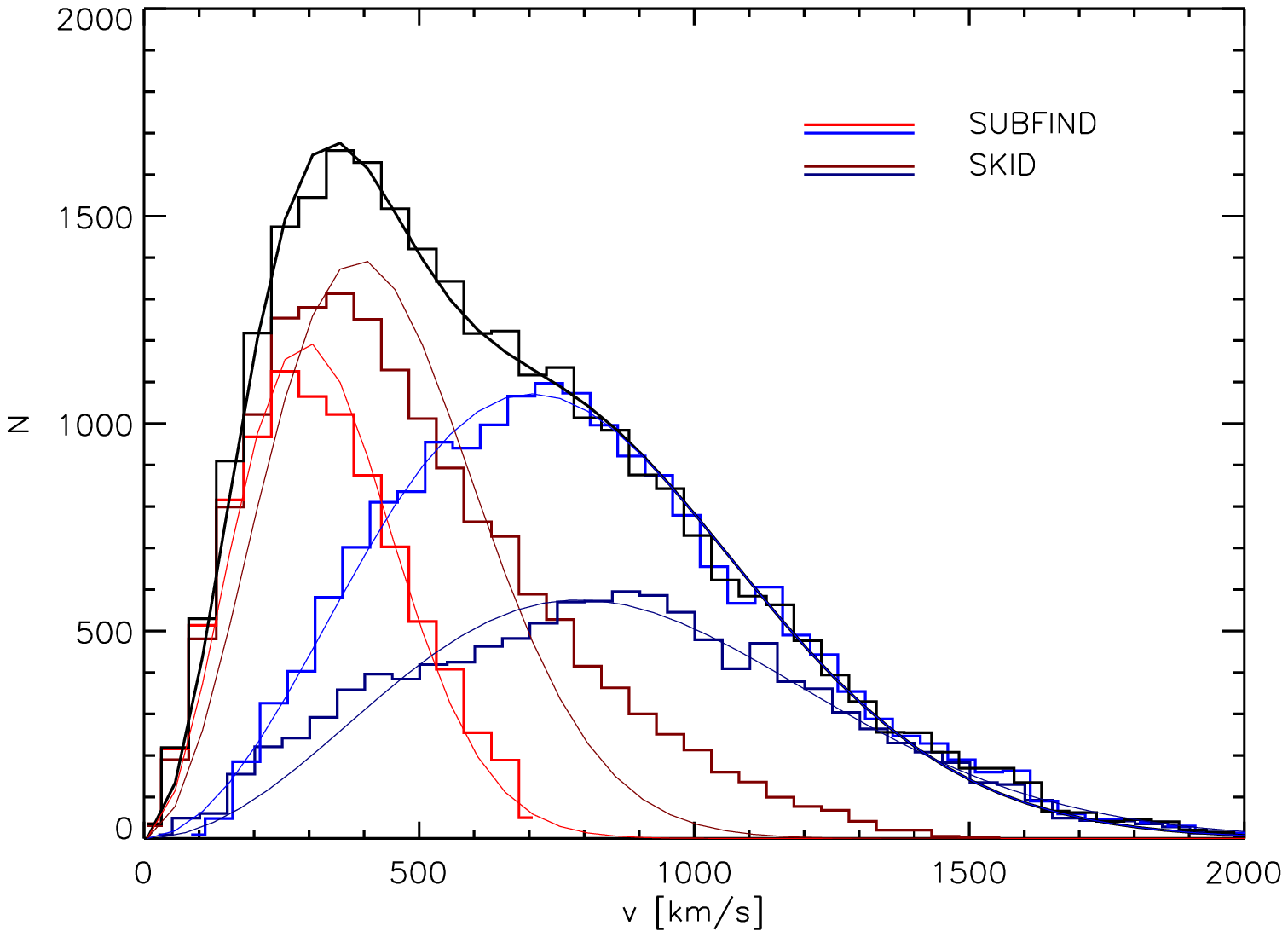}
\caption{As in Figure \ref{fig:comp_3}, we show the velocity histogram
  (black) of the main halo stellar component and a double Maxwellian
  fit to it (thin line). Additionally, the red and blue histograms
  show the velocity distribution of the {\it cD} and {\it DSC}
  components and the thin lines indicate the individual, Maxwellian
  distributions obtained from the global fit. Light and dark colored 
  histograms are the results obtained with  {\small SKID} and {\small
    SUBFIND} respectively. The thin lines are individual Maxwellian
  fits to the  {\it cD} and {\it DSC} components identified by {\small
    SKID}.  
} \label{fig:comp_5}
\end{figure}

Analyzing a smaller cluster ($M_{vir} \approx 2\times10^{14}M_\odot/h$) from
our simulations reveals the differences between the two algorithms for such
less massive systems. Although the individual satellite galaxies match as
nicely as before, {\small SKID} assigns much more stars to the central {\it
  cD} than {\small SUBFIND} leading to the contrary behavior for the {\it
  DSC} component, which is clearly visible in figure \ref{fig:comp_4}, which, 
similarly to figure \ref{fig:comp_1}, shows a particle by particle comparison 
between the two algorithms. The differences in disentangling the {\it cD}
and {\it DSC} component is also clear from figure \ref{fig:comp_5}, which shows the
velocity distribution histograms of the inferred {\it cD} and {\it DSC}
components. Whereas the {\small SUBFIND} results follow nicely, by construction,
the individual components of the double Maxwellian fit to the global
velocity distribution, the {\small SKID} results deviate significantly. In
addition, we fitted individual Maxwellian distributions to the {\small SKID}
results, shown as thin lines. It appears quite obvious that even such fits are
not as good as the ones obtained with our new {\small SUBFIND}
algorithm. Especially the {\it cD} component identified by {\small SKID} shows
a significant tail towards higher velocities which can't be represented by a
single Maxwellian distribution, indicating that the star particles associated
to the {\it cD} by {\small SKID} are not a single, dynamical component.

To obtain better result with {\small SKID}, one should probably tune the $\tau$
parameter on a cluster-by-cluster basis. Such tuning is not needed with our
new {\small SUBFIND} scheme.

\end{document}